\documentclass[preprintnumbers, twocolumn,showkeys,superscriptaddress,showpacs,
nofootinbib,byrevtex,fleqn,prd,notitlepage]{revtex4-1}
\usepackage{bm,pifont}
\usepackage{amssymb}
\usepackage{amsmath}
\usepackage{amsfonts}
\usepackage{epsfig}
\usepackage{graphicx}
\usepackage{tabularx}
\usepackage{color}
\usepackage{slashed}
\newcommand{\seq}{\begin{subequations}}
\newcommand{\sen}{\end{subequations}}
\newcommand{\eq}{\begin{eqnarray}}
\newcommand{\en}{\end{eqnarray}}

\usepackage{lipsum}

\def\shiftdown#1{#1\llap{\lower.04ex\hbox{#1}}}

\def\nn{\nonumber}

\begin{document}
%\title{ ALP  portal: SM and Dark photons mixing from fixed target experiments}
%\title{Probing dark axion portal with fermion EDM and  NA64$e$, NA64$\mu$, LDMX and $\mbox{M}^3$.}	
\title{Passage of millicharged particles in the electron beam-dump: refining constraints from SLACmQ and estimating sensitivity of NA64$e$.}	
\date{\today}
	
\author{Nataliya~Arefyeva \footnote{{\bf e-mail}: arefnat8@gmail.com}}
\affiliation{Physical Department, Lomonosov Moscow State University,
   Vorobiovy Gory, Moscow 119991, Russia} 
\affiliation{Institute for Nuclear Research of the Russian Academy 
	of Sciences, 117312 Moscow, Russia} 
\author{Sergei Gninenko\footnote{{\bf e-mail}: sergei.gninenko@cern.ch}}
\affiliation{Institute for Nuclear Research of the Russian Academy 
	of Sciences, 117312 Moscow, Russia}  
	\author{Dmitry Gorbunov\footnote{{\bf e-mail}: gorby@ms2.inr.ac.ru}}
\affiliation{Institute for Nuclear Research of the Russian Academy 
	of Sciences, 117312 Moscow, Russia} 
	\affiliation{Moscow Institute of Physics and Technology, 
	Institutsky lane 9, Dolgoprudny, Moscow region, 141700, Russia} 
\author{Dmitry~Kirpichnikov \footnote{{\bf e-mail}: kirpich@ms2.inr.ac.ru}}
\affiliation{Institute for Nuclear Research of the Russian Academy 
	of Sciences, 117312 Moscow, Russia}

%\begin{abstract}
%Mixing of Standard model (SM) photon and dark photon can be connected trough axion portal to dark matter. In 
%particular, axion-like particles (ALPs) can be connect SM photon with dark photon. Minimal, leptofilic and hadrofilic 
%scenaries of ALPs are considered for estimation of interaction couplings with dark matter. Couplings will be limited 
%by estimation of invisible mode search in experiments at fixed target.     
%\end{abstract}

\begin{abstract}
Millicharged particles (MCPs) arise in many well-motivated extensions of the Standard Model and are a popular subject for 
experimental searches.  We investigate attenuation of the MCP flux produced at accelerator experiments due to  their interactions in  the media.  
Considering, as an example,  the dedicated MCP search at SLACmQ, we  demonstrate  that this effect can
significantly  affect the final sensitivity to the MCP parameter space leaving its essential part still
unexplored. Applying our analysis to the SLACmQ experiment\,\cite{Prinz:1998ua}, we
correct their exclusion bounds in close accordance with Ref.\,\cite{Prinz:2001qz}.
We also show that  this newly reopened area  with the MCP masses in the  range  
$10^{-4}$\,eV - $1$\,GeV and charges  $ \gtrsim 10^{-5} e$   
can be effectively probed by the  NA64$e$ experiment at the CERN SPS. Light MCPs are mostly produced by virtual photon in electron scattering off nucleus. The main source of heavy MCP is decays of vector mesons, produced by the electrons on nuclei. 
\end{abstract}

\maketitle

\section{Introduction}

The millicharged particles (MCPs) are considered in connection with electric charge quantization mechanism 
and new physics models with electric charge non-conservation~\cite{Ignatiev:1978xj}.  Moreover,
scenarios beyond the Standard Model (SM) of particle physics can naturally include particles, which electric charge is a small fraction of the electron charge, 
$Q_{\chi} = e \epsilon \ll e$. In particular, the millicharged particles can be considered as a well-motivated 
dark-matter candidate\,\cite{Brahm:1989jh,Pospelov:2007mp,Feng:2009mn,Cline:2012is,Tulin:2012wi,Liu:2019knx,Creque-Sarbinowski:2019mcm}. Thus, experimental searches for 
such particles are of a great interest~\cite{Harnik:2019zee}.  

The simplest way to introduce the  MCPs in a 
model is to consider them as a low-energy limit of the theory, where a
hidden (dark) photon, $A_\mu'$, 
kinetically mixes with the visible SM photon~$A_{\mu}$\,\cite{Holdom:1985ag}. As a result, e.g. a new fermion of the hidden 
sector $\chi$, coupled to the hidden photon, can acquire a small electric charge $\sim e\epsilon$, and the model Lagrangian can be written as follows
\begin{equation}
\mathcal{L} \supset i \bar{\chi} \gamma^\mu \partial_\mu \chi - m_{\chi} \bar{\chi} \chi +
e \epsilon A_\mu \bar{\chi} \gamma^\mu  \chi,    
\end{equation}
where  $m_{\chi}$ is the Dirac mass of the hidden MCP.  

The parameter space of the MCP on 
$(\epsilon, m_\chi)$ plane can be constrained from the searches at collider~\cite{Izaguirre:2015eya,Liu:2018jdi,Zhang:2019wnz,Liu:2019ogn,Liang:2019zkb,Bai:2021nai,Ball:2020dnx,milliQan:2021lne,Jaeckel:2012yz} and  fixed-target ~\cite{Chu:2020ysb,Kelly:2018brz,Kim:2021eix,Gorbunov:2021jog,Marocco:2020dqu,Magill:2018tbb,ArgoNeuT:2019ckq,Foroughi-Abari:2020qar,Berlin:2018bsc} experiments, with optical sensors~\cite{Afek:2020lek,Gabrielli:2016rhy,Moore:2014yba} and searches with a superconducting 
radio-frequency cavity~\cite{Berlin:2020pey}, from analysis of cosmological and
astrophysical observations~\cite{Harnik:2020ugb,Dolgov:2016gxu,Li:2020wyl,Aboubrahim:2021ohe,Davidson:2000hf,Vogel:2013raa,Caputo:2019tms,Korwar:2017dio,Ejlli:2017uli,Huang:2015gta,Dolgov:2013una,Melchiorri:2007sq,Dubovsky:2003yn}, and by using results of cosmic-ray detectors~\cite{ArguellesDelgado:2021lek,Kachelriess:2021man,Plestid:2020kdm} and 
nuclear reactor experiments~\cite{TEXONO:2018nir,Chen:2014dsa,Gninenko:2006fi}, etc. 

%New physical scenarios extending the bounds of the Standard Model includes the existence of particles with a small electric charge ($\epsilon e \ll e$), so-called millicharged particles (MCP) \cite{Magill:2018tbb}.
%Although the Standard Model (SM) does not impose the principle of charge quantization, beyond its frames several models were proposed in which charge quantization may be violated, and particles with a small fractional charge appear as a natural consequence of many of these mechanisms \cite{Prinz:1998ua}. 
%The absence of evidence of the fundamental principles of quantization continues to motivate the MCP searches. In their simplest form, MCP can be introduced as new particles violating charge quantization in the SM. The more elegant introduction of MCP is to consider them as a low-energy limit in the theory, where a new dark photon kinetically mixes with the visible; as a result the new particles of hidden sector ($\chi$) can acquire a small electric charge, which can be expected to detect \cite{Harnik:2019zee} 
%$$\mathcal{L}\supset \epsilon e A^\mu \overline{\chi}\gamma^\mu\chi.$$ 
%Millicharged particles may also form the part of dark matter in the Universe. In recent years, %MCP are widely discussed in publications containing various constraints on their mass ($m_\chi$) and charge ($\epsilon$) parameters obtained in laboratory experiments and experiments on accelerators, as well as from astrophysical and cosmological considerations \cite{Magill:2018tbb,Harnik:2019zee}.

In accelerator experiments, the MCPs can be produced at intense proton fixed-target facilities
in decays of mesons 
\begin{equation}
    \pi^0, \eta, \eta' \to \gamma \chi \bar{\chi}, \qquad \rho, \omega, \phi, 
    J/ \psi \to \chi \bar{\chi},   
\end{equation}
from hadronic showers initiated by high-energy protons in a dump.  A thick  shield between the production  and detection points of the MCP absorbs all  strongly and electromagnetically interacting secondary particles, while the MCP may pass through the shield and can be detected in a far detector via  their scattering off electrons,   $\chi e^- \to \chi e^-$. In this case the observation  of MCP  is based on a search for excess  of low-energy recoil electrons 
 in the detector. The number of signal events $N_{\chi}$ in such type of experiments,  scales as 
$N_{\chi} \propto \epsilon^4$ for a single MCP hit in the detector, and as $N_{\chi}\propto 
\epsilon^6$ for the double-hit  MCP signature~\cite{Gorbunov:2021jog}. Consequently, a  large number of protons on target, $\mbox{POT}$,  is required to probe 
small values of $\epsilon$. E.g.,  for, say, $\epsilon \lesssim 10^{-3}$ and MCP masses $m_{\chi} \gtrsim 100$~MeV, one needs    
$\sim \mathcal{O}(10^{20})-\mathcal{O}(10^{22})~\mbox{POT}$.

\par A more powerful and effective approach in probing MCP can be based on their production  in  high energy electrons scattering off heavy nuclei, 
%\[eN\to e N \chi \bar{\chi}\,,\]
the process exploited by the NA64$e$ experiment in the North Area of the CERN SPS. 
The  NA64$e$ facility has been originally designed for  searching  for Light Dark Matter (LDM) production in invisible decays of a dark photon ($A'$) mediator  
 in the reaction chain, $e^- N \to e^- N A'; A'\to {\rm LDM}$ ~\cite{Gninenko:2018ter}. That approach can be also adopted for searching  for MCP  in missing energy events due to the MCP  production in the reaction  $eN\to e^-N \gamma^* \to e^- N \chi \bar{\chi}$ accompanied by  missing energy carried away by the MCP pair. The expected number of signal events in this case  is proportional to $N_\chi \propto \epsilon^2$. Therefore,  a much lower number of electrons 
 on target (EOT),   $n_{EOT}\simeq \mathcal{O}(10^{12})-\mathcal{O}(10^{13})$ is required to test the same region of
 at small values  $\epsilon\lesssim 10^{-3}$\,\cite{Gninenko:2018ter} as compared to the required number of POT.  
 %In case of NA64$e$, one may argue that the interesting statistics of electrons on target starts from $\mbox{EOT}\propto \mathcal{O}(10^{12})-\mathcal{O}(10^{13})$\,    
%$\mbox{POT}\propto \mathcal{O}(10^{20})-\mathcal{O}(10^{22})$.  
\par Another point, discussed in this work, is related to the attenuation of a charged particle flux while propagating in matter. The  energy losses and multiple scattering of charged particles  strongly depend on the their charge to mass ratio, as we well know from the comparison of the electron and muon propagation in medua. Therefore,  
for the relatively light  MCP in the mass range, $m_\chi \lesssim 10$~MeV,   and a small charge $\epsilon \lesssim 10^{-5}-10^{-1}$,  the search sensitivity of  
 beam-dump experiments, such, e.g. as SLACmQ~\cite{Prinz:1998ua}, might by constrained
 by a significant attenuation of the MCP flux at the detector location.  
  \par In this paper we reconsider part of the MCP bounds  from  
the SLACmQ beam-dump experiment  taking into account the MCP flux attenuation due to i) MCP energy loss in the reactions of  $\chi e^- \to \chi e^-$ scattering, pair production, $\chi N\to \chi N e^+ e ^-$, bremsstrahlung, $\chi N \to \chi N \gamma$ and ii) decreasing of the MCP deflection acceptance due to the MCP multiple scattering in the dump. We show that indeed,  altogether these processes may prevent the MCP from reaching the SLAC mQ detector and giving the signal.    
We find that our revised results on MCP sensitivity are in a fairly good agreement with previous study~\cite{Prinz:2001qz}. We implement the developed  
ideas of MCP stopping power calculation for the estimate of expected reach of NA64$e$ in the region of model parameter space,  
$m_{\chi} \lesssim 100$\,MeV and $\epsilon \lesssim 10^{-5}-10^{-1}$. We also calculate and find it promising, the expected sensitivity of 
NA64$e$ to MCP, produced in bremsstralung-like events, $eN \to e N \gamma^* (\to \chi \bar{\chi}) $ and to MCP, emerged 
in invisible decays of short-lived vector mesons $V=\{ \rho, \omega, \phi, 
    J/ \psi\} $, produced in the reaction of photo-production inside the dump,  
$\gamma^* N \to N V  (\to \chi \bar{\chi})$ for $m_{\chi}\gtrsim 10$~MeV and $10^{-3} \lesssim \epsilon \lesssim 10^{-2}$.

%Since it is 
%believed that MCP participate only in electromagnetic interaction, their passage through the matter depends on the charge. For this reason, even a relatively weakly interacting particle lose a significant fraction of its energy when pass the matter and the detected signal will be minor. Thus, with the increase in the sensitivity of experiments to low-energy signals, there is hope of finding these particles.

%The aim of the paper is to study the passage of hypothesized MCP through a matter and try to determine the area of 
%insensitivity. In this paper the propagation of MCP thought matter and their energy loses are calculated. The new 
%look to the parametric mass-charge space that consider the matter effect is described and the new constrains on MCP
%using data of the main existing experiments searching MCP are obtained.

The paper is structured as follows. Section~\ref{DerivStopLossAndScatt} describes 
various effects associated with MCP passage through matter, namely: the dominant MCP energy  losses in matter 
(ionization and radiation losses via bremsstrahlung and $e^+e^-$ pair production) and  the deflection of MCP trajectory due to multiple scattering. In Section~\ref{RevisSLACmQ} we implement the results of 
Section~\ref{DerivStopLossAndScatt} to revise the analysis of the SLACmQ experiment\,\cite{Prinz:1998ua} and correct their exclusion region 
in $(m_\chi,\epsilon)$ parameter space. In Section~\ref{NA64Cross-section} we calculate the exact tree-level 
cross section of the process $eN \to e N  \gamma^*(\to \chi \bar{\chi})$ to estimate the yield
of the MCP in the NA64$e$ missing energy signature.  In 
Section~\ref{NA64SensitivitySection} 
we estimate the NA64$e$ prospects in probing MCP models for the relatively wide mass range
$10^{-4}\, \mbox{eV} \lesssim m_\chi \lesssim 1~\mbox{GeV}$. In 
Section~\ref{VectorMesonReach} we estimate the expected reach of NA64$e$ to examine MCPs 
from invisible decay of vector mesons.  We conclude in Section\,\ref{Conclusion}. Two appendices contain some formulas used in the numerical calculations.

%\section{Kirpichnikov: addendum (alternative) to introduction}

%%%%%%%%%%%%%%%%%%%%%%%%%%%%%%%%%%%%%%%%%%%%%%%%%%%%%%%%%%%%%%%%%%%%%%%%%%%%%%%
\section{MCP interactions in  matter
\label{DerivStopLossAndScatt}}

\begin{figure*}[!htb]
\centering
\label{fig:my_label-4}
\includegraphics[width=\textwidth]{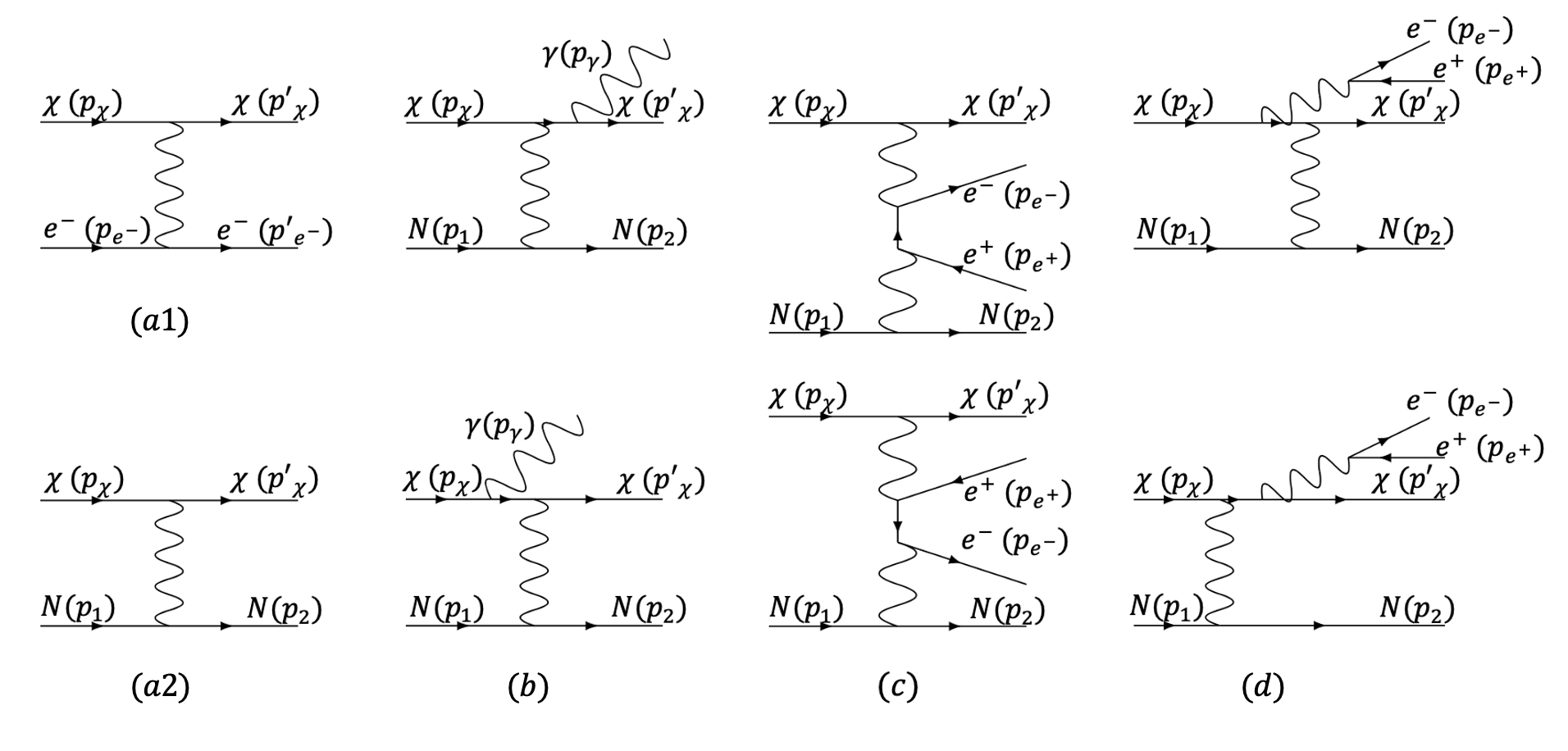}
\caption{Feynman diagrams for MCP energy loss in matter: (a) scattering off atomic nucleus and electrons, (b) bremsstrahlung, (c,d) $e^+e^-$ pair production.  
\label{Feynmandiag} }
\end{figure*}

\qquad High energy particles can produce sufficiently light MCPs by scattering off target. In turn, the 
MCP collides with electrons and nuclei of the dump material resulting in energy losses and
multiple scattering, and thus, deviation from its original direction, when passing through  matter. In this Section we consider 
three processes of MCP energy loses: ionization, radiation and pair production, some of which are illustrated with Fig.\,\ref{Feynmandiag}. 
In the end of Section we estimate a typical angle of MCP deviation from the original
direction due to the  multiple scattering.
%All the expressions are given in a natural system $\hbar=c=1$ unless otherwise specified.  
% \begin{figure}[!tbh]
% \centering
% \includegraphics[width=0.495\textwidth]{ioniz.png}
% \includegraphics[width=0.495\textwidth]{brems.png}
% \includegraphics[width=0.495\textwidth]{pair.png}
% \caption{Feynman diagrams for MCP energy loss in matter, (a) scattering from atomic nucleus and electrons, (b) bremsstrahlung, (c) $e^+e^-$ pair production. 
% \label{Feynmandiag}
% }
% \end{figure}

%%%%%%%%%%%%%%%%%%%%%%%%%%%%%%%%%%
\subsection{Ionization losses}

% \qquad  Collisions of a relativistic MCP with an atom accompanied
% by its excitation (discrete spectrum of the atom final state) or ionization (continuous spectrum)
% are added to the MCP ionization energy losses (Figure ~\ref{Feynmandiag} (a)).
 Ionization energy losses are associated with the MCP collision with an atom, which initiates the atom 
excitation and knocking out the electron, see Fig.\,\ref{Feynmandiag}\,(a1).
To quantify the ionization losses effect for MCP we adopt the Bethe--Bloch formula as 
% an
% integrated expression of (\ref{eq1}) was used 
\cite{ParticleDataGroup:2020ssz}:
%\begin{equation}\label{eq3} 
%\left(-\dfrac{dE_\chi}{dx}\right)_{\text{ion}}=\dfrac{4\pi Z \epsilon^2 r_e^2 m_e }{\beta_\chi^2}\, \dfrac{N_A\rho}{M_A}\left(\dfrac{1}{2}\ln{\dfrac{2m_e \beta_\chi^2 \gamma_\chi^2 T_{max}}{I^2}} - \beta_\chi^2 - \dfrac{\delta(\beta_\chi\gamma_\chi)}{2}\right)
%\end{equation}
\begin{align}
& \left(-\dfrac{dE_\chi}{dx}\right)_{\text{ion}}= \dfrac{4\pi Z \epsilon^2 r_e^2 m_e }{\beta_\chi^2}\, \dfrac{N_A\rho}{M_A} \times \nn
\\
& \times \left(\dfrac{1}{2}\ln{\dfrac{2m_e \beta_\chi^2 \gamma_\chi^2 T_{max}}{I^2}} - \beta_\chi^2 - \dfrac{\delta(\beta_\chi\gamma_\chi)}{2}\right)
  \label{eq3} 
\end{align}
where $x$ is the track length of MCP,  $\beta_\chi$ is its velocity and $\gamma_\chi=E_\chi/m_\chi $ is its Lorentz factor.  The last term 
in \eqref{eq3} comes from the matter polarization, and at very high energies, $\gamma_\chi\gg1$, it approaches   $\delta(\beta_\chi\gamma_\chi)\approx 2\log(\omega_p\beta_\chi\gamma_\chi/I)-1$. Then  $I$ stands for the ionization potential of the atom,
% connected with the fact that at $\gamma_\chi\gg1$ the maximum
% value of the impact parameter $\hbar c \gamma_\chi/I$ can become much larger than the
% distance between the matter atoms, and the field of the particle is noticeably less than а
% Coulomb field. Thus the increase in the impact parameter with MCP energy growth will stop,
% the plasma energy is $\hbar \omega_p=22.8\sqrt{\rho Z/A}$ eV; 
 $r_e=\alpha/m_e \simeq 2.8\, \mbox{fm}$ is the classical radius of electron, $N_A=6.02\cdot 10^{23}$mol$^{-1}$ is the Avogadro's number, $Z$ is the atomic number of target material, $M_A$ is the atomic mass of the target material and
 $\rho$ is its density, the maximum energy transferred from MCP to electron is given by
\begin{equation}\label{eq2} 
\begin{gathered}
%m_e c^2\, \Delta_{max}=\dfrac{2m_e c^2\,|\vec{p_\chi}|^2 c^2}{m_\chi^2c^4+m_e^2c^4+2m_e c^2 E_\chi}=\\
T_{max} =\dfrac{2m_e\beta_\chi^2\gamma_\chi^2}{1 + 2\gamma_\chi m_e/m_\chi + m_e^2/m_\chi^2}.
\end{gathered} 
\end{equation} 
We note,  that the MCP's ionization stopping power (\ref{eq3}) scales as $\propto \epsilon^2$ and 
depends logarithmic on $E_\chi$ and $m_\chi$ in the ultra-relativistic regime $\beta_\chi\simeq 1$.

%%%%%%%%%%%%%%%%%%%%%%%%%%%%%%%%%%%%%%% 
\subsection{Radiation losses}

The MCP slow down due to scattering off nuclei in the material and emission of 
photons as illustrated by the Feynman diagrams in 
Fig.\,\ref{Feynmandiag}\,(b). 
The radiation loss due to scattering off electrons is suppressed by the nuclear charge as $1/Z$, and we neglect it in our study.

The atomic electrons outside the nucleus screen the nucleus  
Coulomb field, so that its effective strength decreases and atomic electrons also serve as scattering targets. In our case of ultra relativistic MCP the screening can be considered as a complete one. %($F_{corr} = [Z^2(L_{rad}-f(Z))+ZL_{rad}']+\dfrac{1}{9}(1-\dfrac{E_\gamma}{E_\chi})(Z^2+Z)$)
The MCP energy losses for radiation per unit length is determined by 
\begin{equation}\label{eq4} 
\left(-\dfrac{dE_\chi}{dx}\right)_{\text{brems}}=n\,\int\limits_0^{E_\chi-m_\chi} dE_\gamma   \frac{d\sigma_{\text{brems}}}{dE_\gamma}  E_\gamma\,,
\end{equation}
where $n=N_A \rho/M_A$ is the number density of atoms in the target, $d\sigma_{\text{brems}}/d E_\gamma$  is the differential 
bremsstrahlung cross section, $E_\gamma$ is the energy of the emitted photon.
The bremsstrahlung spectrum in the case of complete screening is given by the formula \cite{ParticleDataGroup:2020ssz} 
\begin{equation}\label{eq6} 
\frac{d\sigma_{\text{brems}}}{dE_\gamma} \simeq \dfrac{1}{E_\gamma}\,4\alpha Z^2  r_e^2\,\epsilon^4\!\!\left(\dfrac{m_e}{m_\chi}\right)^2 \!\!
\left( \dfrac{4}{3}-\dfrac{4}{3}\dfrac{E_\gamma}{E_\chi}+\dfrac{E_\gamma^2}{E_\chi^2} \right)
F_{corr}\,, 
\end{equation}
where $F_{corr}$ is the factor which takes into account the Coulomb correction to the Born cross section and 
atomic screening effects \cite{ParticleDataGroup:2020ssz}.
We note that the stopping loss of the MCP due to the radiation scales as $\propto \epsilon^4 E_\chi /m_\chi^2$, therefore, the lighter MCPs loose their  energy more rapidly.

%Accounting the MCP radiation, the differential cross section takes the term $\dfrac{\epsilon^4 m_e^2}{m_\chi^2}$.
%\noindent $f(Z)=(\alpha Z)^2\sum\limits_{n=1}^{\infty}\dfrac{1}{n(n^2+(\alpha Z)^2)}$ is the Coulomb correction to the Born cross section; the terms accounting for the screening are: $L_{\text{rad}}=\ln{(184.15\, Z^{-\frac{1}{3}})}$, $L_{\text{rad}}'=\ln{(1194\, Z^{-\frac{2}{3}})}$

%%%%%%%%%%%%%%%%%%%%%%%%%%%%%%%%%%%%%

\subsection{Electron-positron pair production} 

\begin{figure*}[!htb]
\centering
\includegraphics[width=0.495\textwidth]{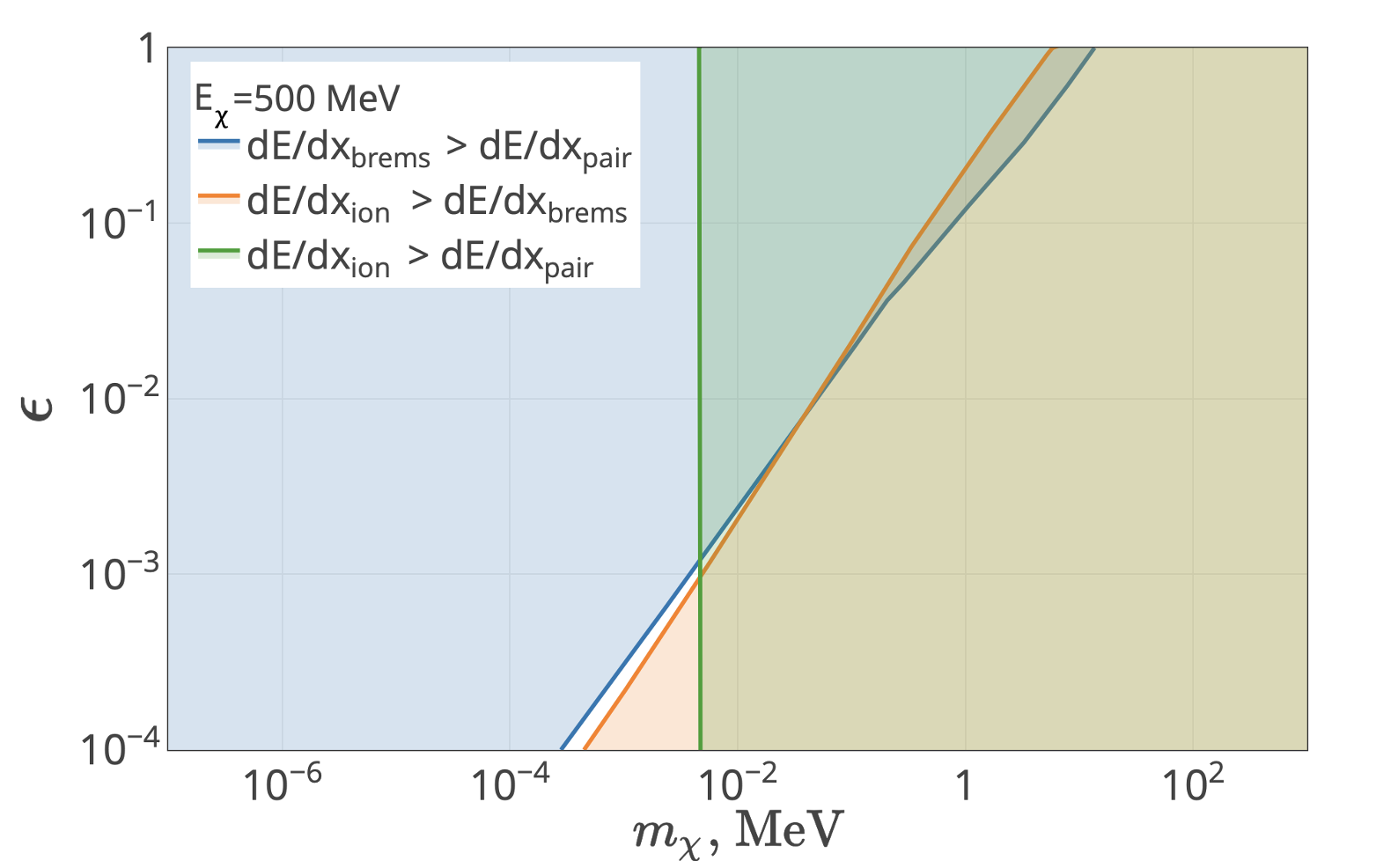}
\includegraphics[width=0.495\textwidth]{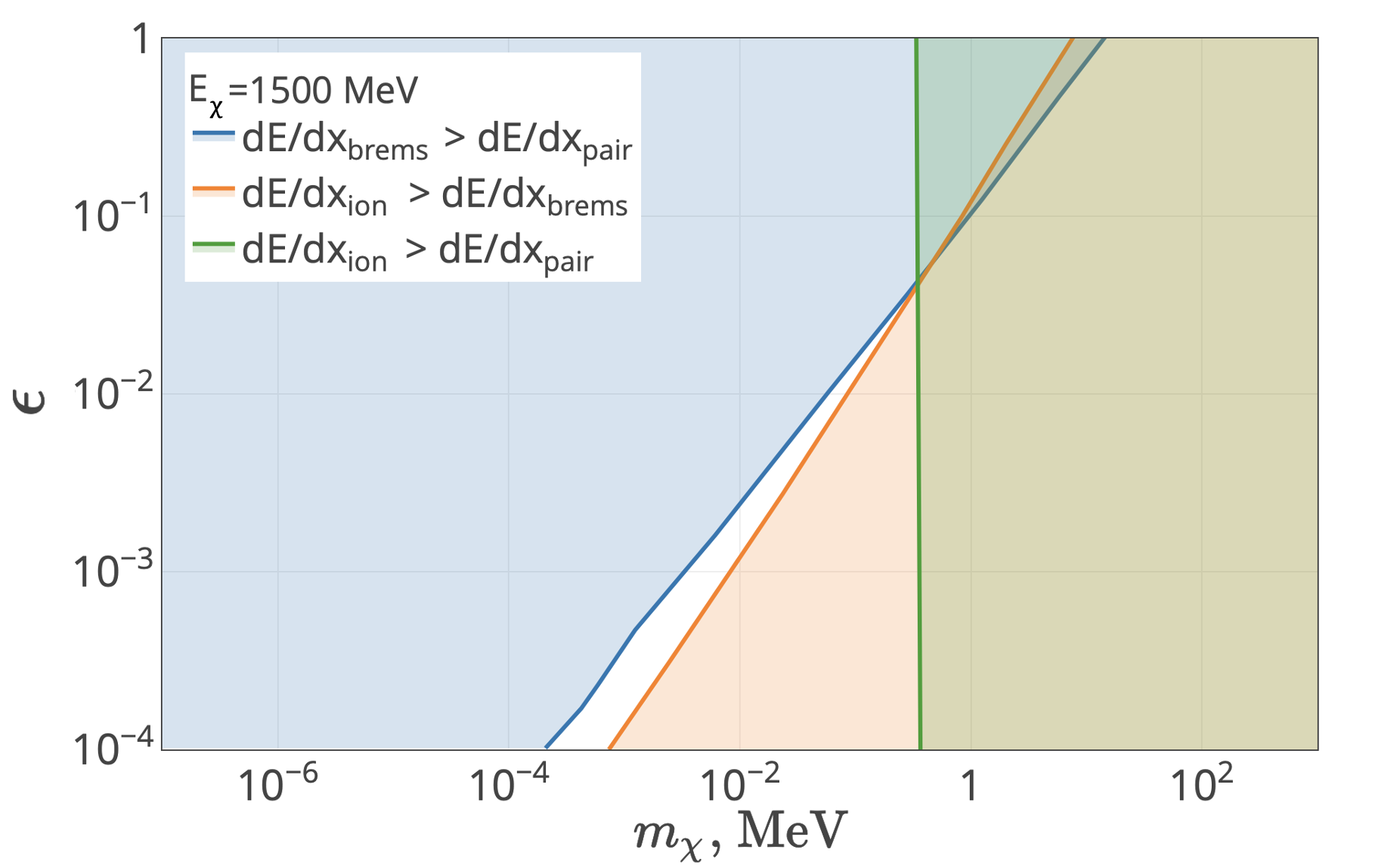}
\includegraphics[width=0.495\textwidth]{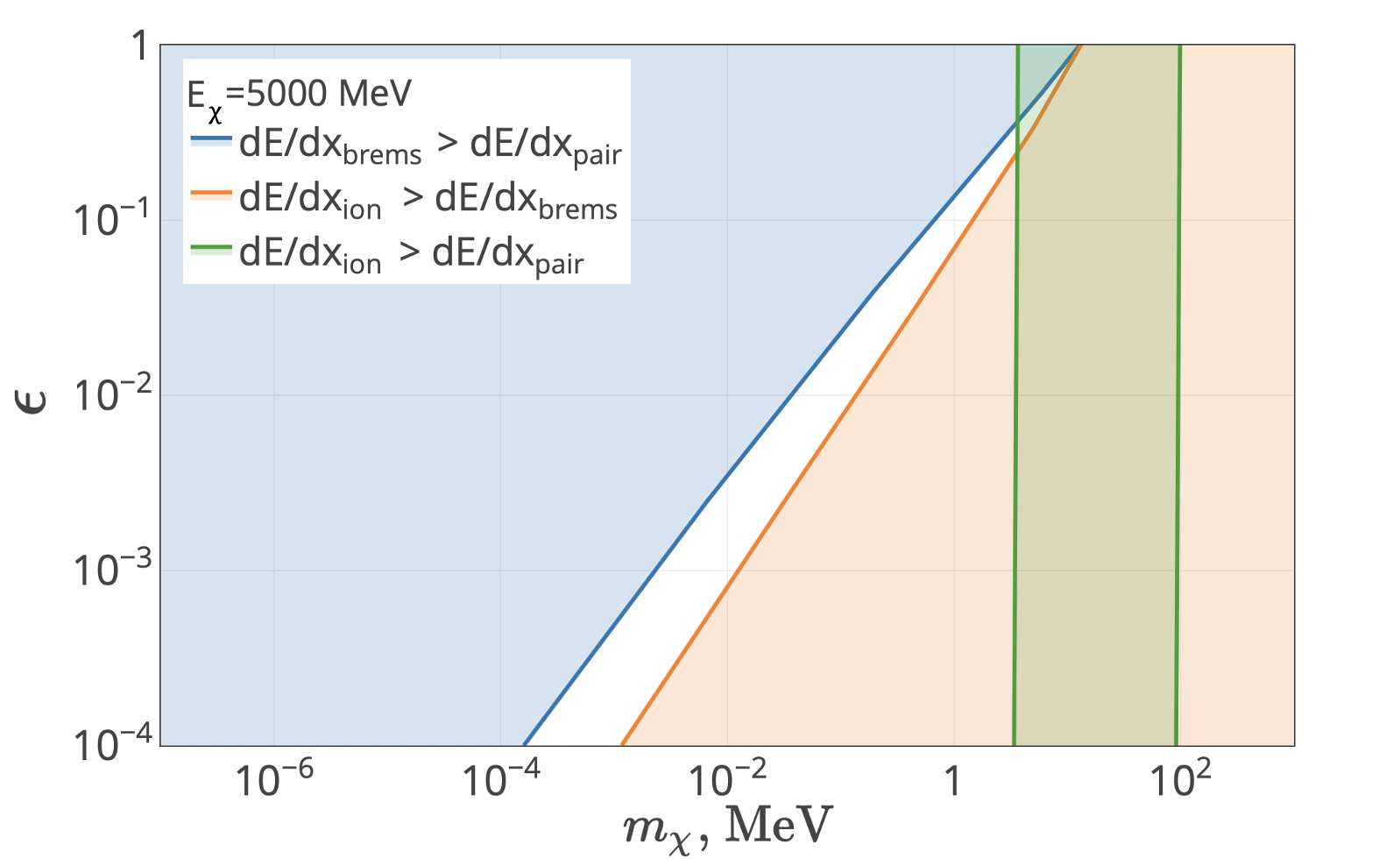}
\includegraphics[width=0.495\textwidth]{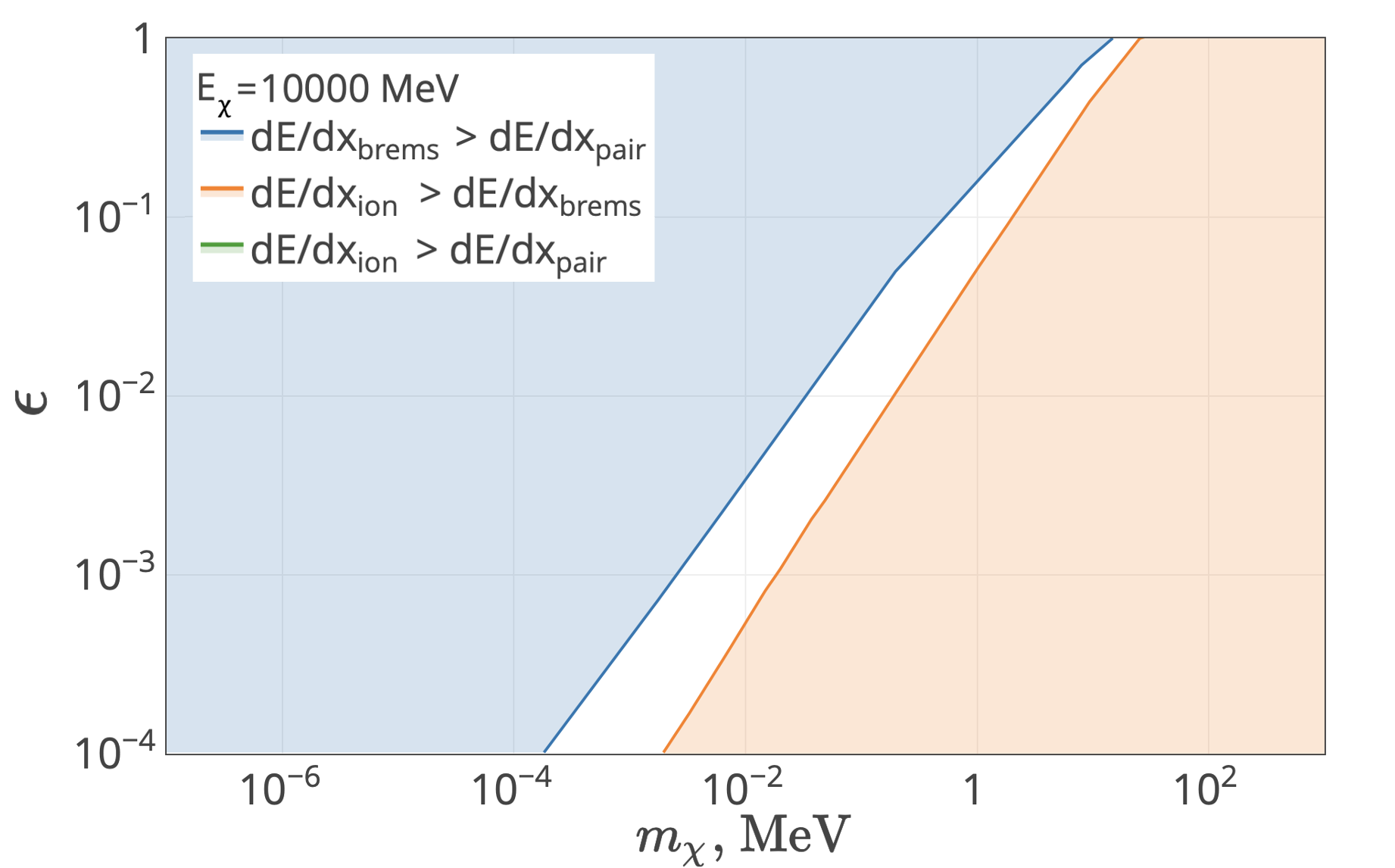}
\caption{The diagrams showing in which parts of the model parameter space, which  mechanism of the energy loss dominates. There are four plots for MCP energies  $E_\chi= 500$\,MeV, $E_\chi= 1.5$\,GeV, $E_\chi=5$\,GeV, $E_\chi=10$\,GeV.  
\label{dominatelosses}
}
\end{figure*}

 MCP energy losses can be accounted by considering the process illustrated with Fig.\,\ref{Feynmandiag} (c), (d). 
 One observes, that for sufficiently small parameter $\epsilon$, the amplitude of Fig.\,\ref{Feynmandiag} (d) is 
 suppressed due to additional factor $ \epsilon $. The MCP's energy losses  due to the $e^+e^-$ pair production 
 can be written as 
\begin{equation}\label{eq9} 
\left(-\dfrac{dE_\chi}{dx}\right)_{e^+e^-\text{ pair}}=n\,\int\limits_{2m_e}^{E_\chi-m_\chi}d\omega\, \dfrac{d\sigma_{e^+e^-\text{ pair}} }{d\omega} \, \omega\,,
\end{equation}
where $\omega=E_{e^+}+E_{e^-}$ is the total energy of $e^+e^-$ pair.  The cross section $d \sigma_{e^+e^-}/d\omega$ can be taken~\cite{Bayer:1973tx} in the 
ultra-relativistic approximation, when the energies of initial and final MCPs are high, $E_\chi',E_\chi\gg m_\chi, m_e$, as well as the energies of electron and positron,  $E_{e^+},E_{e^-}\gg m_e$.
In particular, the integration over the energy of positron yields the spectrum of the produced pair, 
\[
\dfrac{d\sigma_{e^+e^-\text{ pair}}}{d\omega}
=\int_{m_e}^{\omega-m_e} \dfrac{d\sigma_{e^+e^-\text{ pair}}}{dE_{e^+} dE_{e^-}}dE_{e^+}\,,
\]
where the double differential cross section is written as 
\begin{equation}\label{eq7} 
\frac{d\sigma_{e^+e^-\text{ pair}}}{dE_{e^+} dE_{e^-}}=Z^2 \epsilon^2 \, \dfrac{\alpha^2 r_e^2}{2\pi} \, \dfrac{E_\chi'-\omega}{\omega^2 E_\chi} \, G\,, 
\end{equation}
and the value of $G$ is given in Appendix~\ref{appendixA}.
% \begin{equation}\label{eq8} 
% \begin{gathered}
% G = \left[ 2\ln{\dfrac{E_{e^+}\, E_{e^-}}{2m_e\omega\sqrt{1+\xi}}}-1\right]\times\left[ A\, \ln{(1+\dfrac{1}{\xi})}+B+\dfrac{C}{1+\xi}\right]-\\
% -A\, g(\dfrac{1}{1+\xi})-B\xi\ln{(1+\dfrac{1}{\xi})}-\dfrac{C}{1+\xi} 
% \end{gathered}
% \end{equation}
% where 
% \[\xi=\dfrac{m_\chi^2 E_{e^+}\, E_{e^-}}{4m_e^2E_\chi (E_\chi-\omega)}; \quad g(x)=-\int_0^x\ln{\dfrac{|1-t|}{t}}dt;\]\[ A=(1-\dfrac{4}{3}\dfrac{E_{e^+}\, E_{e^-}}{\omega^2})(1+\dfrac{\omega^2}{2 E_\chi E_\chi'})+\dfrac{4}{3}\xi(1-\dfrac{E_{e^+}E_{e^-}}{\omega^2});\] \[B=\dfrac{4}{3}\dfrac{E_{e^+}E_{e^-}}{\omega^2}-1; \quad C=-\dfrac{\xi}{3}-\dfrac{1}{6}\dfrac{\omega^2}{E_\chi E_\chi'}-\dfrac{1}{3}\dfrac{(E_{e^+}-E_{e^-})^2}{\omega^2}\]

In the given material the energy losses of all the types depend on the MCP energy $E_\chi$, mass $m_\chi$ and  parameter $\epsilon$. We illustrate their impacts with plots of Fig.\,\ref{dominatelosses}. 
We observe that in different parts of the model parameter space  different mechanisms dominate at a given energy. In particular, one can neglect the ionization energy loss of the MCP at energies $E_\chi \gtrsim 10$~GeV. 

The energy loss of the MCP on its way from the production point through the media to the detector, can significantly reduce its energy. It should be taken into account when estimating the energy release in the detector due to the MCP elastic scattering inside the detector, which provides the MCP signal signature. In case of large energy loss the MCP energy may drop below the threshold accepted for the energy release to be observable. Such a low-energy MCP avoids the detection.   

%%%%%%%%%%%%%%%%%%%%%%%%%%%%%%%%%%%%%%%%%

\subsection{Multiple scattering}

 We should also take into account the deflection of the MCP trajectory from the rectilinear one, which happens mostly due to multiple elastic scatterings off nuclei and electrons in the media, see the corresponding Feynman diagrams in Fig.\,\ref{Feynmandiag} (a1) and (a2).  
For the nucleus case the averaged squared deflection angle of the MCP per 
unit propagation length reads
\begin{equation}\label{eq11} 
\dfrac{d\overline{\theta}_{\chi Z}^2}{dx}=n\int\limits_0^\pi  d\theta_\chi \,  \theta_\chi^2\, \frac{d\sigma_{\chi Z}}{d\theta_\chi}
\end{equation}
with the scattering cross section on a nucleus 
% \[
% d\sigma_{_Z}=\dfrac{1}{16\pi^2}\,|M_{fi}|_{_Z}^2\sin\theta d\theta d\phi
% \]
% \[
% |M_{fi}|_{_Z}^2=\dfrac{\epsilon^2\,e^2}{2}\left|A_0(\vec{q})\right|^2Tr\left[(m_\chi+\gamma^{\mu}p'_{\chi\mu})\gamma^0(m_\chi+\gamma^{\nu}p_{\chi\nu})\gamma^0\right]
% \]
\begin{equation}\label{eq10} 
\frac{d\sigma_{\chi Z}}{d\theta_\chi}=\dfrac{Z^2\,\epsilon^2\, r_e^2\, m_e^2}{4\, E_\chi^2\, \beta_\chi^2}\dfrac{1-\beta_\chi^2\sin^2\left(\frac{\theta_\chi}{2}\right)}{\sin^2\left(\frac{\theta_\chi}{2}\right)+1/\left(2\, a\, p_\chi\right)^{2}}
\,2\pi\sin\theta_\chi\,,
\end{equation}
where $a \simeq 111 Z^{-1/3}/m_e$ is the screening parameter for the Thomas--Fermi atoms. 

%(large transferred
%momentum $q$: $|q|>\alpha m_e$)). The nucleus is considered motionless $q_0=0, E_\chi=E'_\chi$.
%It is necessary to take into account that the scattering occurs on the shielded potential of the nuclear with the corresponding form factor \cite{litlink10}:
%\[
%A_0(x)=\dfrac{Ze}{4\pi r} \rightarrow A_0(q)\,\dfrac{a^2 q^2}{1+a^2q^2}, \mbox{ $a\approx a_0 Z^{-1/3}$ %is the atom size.}
%\]
%\noindent- for a hydrogen-like atom $a_0=a_B=1/\alpha m_e$ is Bohr radius; 
%\noindent- 

Similarly, for the case of MCP deflection by the atomic electron one has the following 
expression for the average squared angle per unit propagation length
\begin{equation}
\dfrac{d\overline{\theta_{\chi e}^2}}{dx}=Z\,\dfrac{N_A\rho}{M_A}\!\!\int\limits_0^\pi  \theta^2_\chi \, \dfrac{d\sigma_{\chi e}}{d\theta_\chi}\! d\theta_\chi 
\Theta(E_\chi-E_\chi'-\alpha^2 m_e/2)\,, 
\label{eq13} 
\end{equation}
where the differential  cross section in the laboratory frame is 
\begin{align}
    \label{eq12}
&\frac{d\sigma_{\chi e}}{d\theta_\chi} 
=\frac{2}{\pi}  \sin\theta_\chi \dfrac{\epsilon^2 r_e^2 m_e |\vec{p_\chi}'|}{E_e'|\vec{p_\chi}|} \times \\ 
& \times \dfrac{m_e^2(E_\chi^2+E_\chi'^2)+\frac{1}{2}(m_e^2+m_\chi^2)(2m_\chi^2-2(p_{\chi} p_{\chi'}))}{(2m_\chi^2-2(p_{\chi} p_{\chi'}))^2}.  \nn 
\end{align}
The last term in Eq.~(\ref{eq13}) is the step-function, which accounts for screening effects. 
 In particular, the additional condition should be imposed on the MCP transfer momentum squared, 
$-t=-(p_\chi'-p_\chi)^2 \gtrsim 1/R_{scr}^2$, where $R_{scr} \simeq (\alpha m_e)^{-1}$ is the typical screening radius 
for the atomic electrons. The latter condition implies the following inequality  $E_\chi' \lesssim E_\chi-\frac{\alpha^2}{2}m_e$, which defines the step-function $\Theta(x)$ in the integral of \eqref{eq13}. It refers to the fact that the energy of the initial electron is below its mass because of the binding energy $-\alpha^2 m_e/2$. 

%Thus the average deflection angle per unit length take the form:
%\begin{equation}\label{eq13} 
%\dfrac{d\overline{\theta_e^2}}{dx}=Z\,\dfrac{N_A\rho}{M_A}\int_0^\pi %\int^{E_\chi-\frac{\alpha^2}{2}m_e}_{E_\chi-T_{max}}  \theta^2\, \dfrac{d\sigma_e}{dE_\chi'd\theta} %dE_\chi'd\theta 
%\end{equation}

%%%%%%%%%%%%%%%%%%%%%%%%%%%%%%%%%%%%%%%%%%%%%%%%%%%%%%%%%%%%%%%%%%%%%%%

\section{Revisiting SLACmQ bounds on MCPs
\label{RevisSLACmQ}}
 In this Section we revise the SLACmQ bounds on MCPs using 
 detailed information on the expected MCP spectra and geometry of the SLACmQ experiment provided in Ref.~\cite{Prinz:2001qz}. In particular, we reconsider the  limits 
associated with MCP stopping power, time resolution and angular  acceptance of the SLAC detector. The revised constraints we obtain in this way are similar to the ones from 
 Ref.\,\cite{Prinz:2001qz}.  
%%%%%%%%%%%%%%%%%%%%%%%%%%%%%%%%%

\subsection{Stopping power}
\begin{figure*}[!htb]
\centering
\label{fig:my_label-4}
\includegraphics[width=0.85\textwidth]{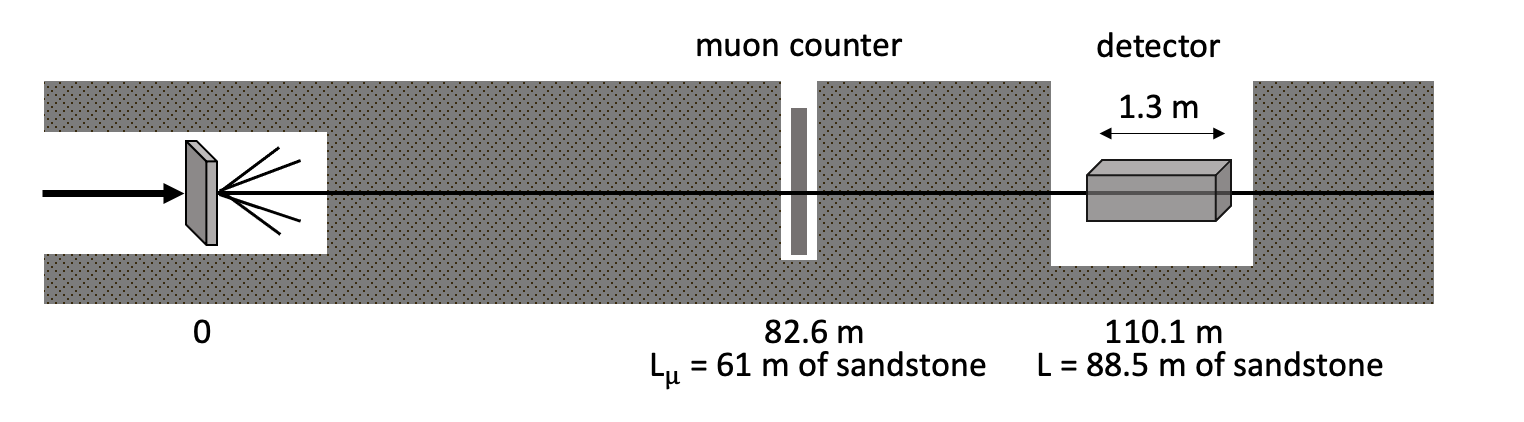}
\caption{Layout of SLACmQ experiment. 
\label{setup SLAC} }
\end{figure*}

Let us consider first the impact of stopping power on bounds on MCP parameter space obtained by the SLACmQ experiment, see its layout in Fig.\,\ref{setup SLAC}. 
The SLACmQ collected data of $3.8\times 10^{18}$ EOT.  
It employed a pulsed beam of $29.5$\,GeV electrons hitting the target made mostly of rhenium ($25\%$) and tungsten ($75\%$). 
On the way to detector, placed at the distance $S=110.1$\,m from the target, MCPs passed downstream  through sandstone, 
 with effective density $\rho=2.19$ g/cm$^3$ and a length of  $L\simeq 88.5$\,m.  Note that MCPs are produced in the target mainly due
to the bremsstrahlung process, $e N\to e N \chi \bar{\chi}$, which implies that  the production 
cross section scales as $\propto \alpha^4\epsilon^2 Z^2$. Therefore, the number of produced MCP pairs 
is proportional to $N_{prod} \propto \epsilon^2$. 

The predominant signature of the MCP used for detection is its elastic
scattering off electrons in the far detector, $\chi e^- \to \chi e^-$, see the corresponding diagrams in Fig.\,\ref{Feynmandiag}(a1). Hence, assuming $\epsilon\ll 1$ and MCP reaching the detector,  the part of MCPs which hit the electrons on the way through the detector scales as $N_{det} \propto \epsilon^2$.  
 In addition, we note that  for $\epsilon \ll 1$ 
and $m_\chi \gg m_e$  both the rate of MCP production and their attenuation rate are small. 
Therefore, as long as the attenuation  rate of MCP is negligible  the  number of expected 
signal  events scales as $N_{sig} \propto \epsilon^4$.  To set the constraints at a given confidence level (typically 90\%C.L. or 95\%C.L.) in that region of MCP's
parameter space one should set $s_{up}\simeq N_{sig}$, where $s_{up}$ is the upper limit on the average 
number of signal events for the given sum of signal and background events.  In Fig.~\ref{Combinedexclusion} these bounds 
correspond to the lower edge of the SLACmQ~\cite{Prinz:2001qz} excluded contour. 
We adopt that curve from Fig.\,5.17  of Ref.\,\cite{Prinz:2001qz}.      
It differs noticeably from the curve presented in the seminal work\,\cite{Prinz:1998ua} as the upper limits of SLACmQ on the parameter $\epsilon$.

Let us consider the attenuation of the MCP initial energy flow $I_0$ 
\[
I=I_0\,\exp{\left(-\dfrac{L}{X_\chi}\right)}\,,
\]
where $X_\chi$ is the MCP attenuation length, $L$ is the distance traveled by the MCP in  matter. In case of the SLACmQ experiment 
the MCP attenuation length can be estimated as follows
\begin{equation}\label{eq15} 
   X_\chi=\int^{E_{\text{max}}}_{E_{\text{min}}} \, dE_\chi \, \frac{1}{N}\dfrac{dN}{dE_\chi} \int^{E_\chi}_{E_{\text{cut}}^\chi}\dfrac{dE_\chi'}{\left|dE_\chi'/dx\right|}\,,
\end{equation}
where $E_\chi$ is initial energy of MCP and $E^\chi_{\text{cut}}$ is the energy threshold of MCP detection  
in the experiment,  $E_{\text{min}}$=0.125 GeV, and $E_{\text{max}}$=29.5 GeV is the beam energy \cite{Prinz:2001qz}.
The total energy losses are:
\begin{equation}\label{eq14} 
    \left(\dfrac{dE_\chi}{dx}\right)_{\text{tot}} = \left(\dfrac{dE_\chi}{dx}\right)_{\text{ion}} + \left(\dfrac{dE_\chi}{dx}\right)_{\text{rad}} + 
    \left(\dfrac{dE_\chi}{dx}\right)_{\text{pair}}, 
\end{equation}
the normalized to unity spectra $N^{-1}dN/dE_\chi$ of the produced MCP are taken from Table A.2 in  Appendix of \cite{Prinz:2001qz}, in Fig.~\ref{Approxspectra} 
\begin{figure}[!htb]
\centering
\label{fig:my_label-4}
\includegraphics[width=0.5\textwidth]{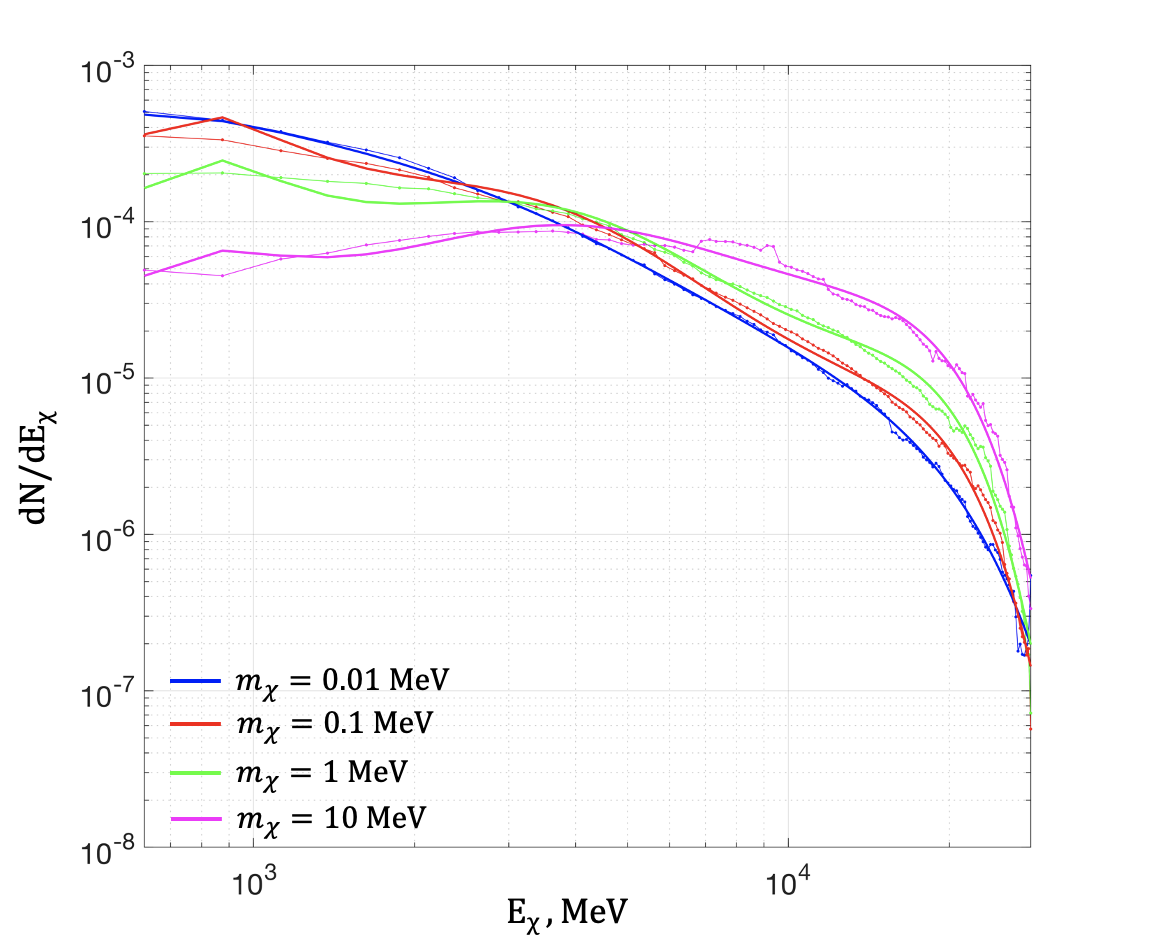}
\caption{Approximations to the differential energy spectra from SLACmQ data  \cite{Prinz:2001qz} normalized to inity for a set of MCP masses $m_{\chi}$ and $\epsilon=10^{-3}$. 
\label{Approxspectra} }
\end{figure}
we show them for various masses $m_\chi$.  
%$\dfrac{dN(E)}{dE_\chi}$ is the differential spectrum show in Figure \ref{Approxspectra}.
%The expression for differential energy spectrum was obtained by the approximation energy distribution by function:
%\[
%F(E)=10^{P_4(\log_{10}E_\chi(m_\chi))}\, \dfrac{1}{N_0}\, ; \quad  %\int^{E_{\text{max}}}_{E_{\text{min}}} F(E)dE = f(m_{\chi})\, 
%\]
%$P_4$ is polynomials of 4th degree, $N_0$ is total number of particles. The normalization %of spectrum to unity takes into account the mass dependence of approximation.  

The account of attenuation of the MCP passing through the matter changes the upper bound on $\epsilon$ at a given mass $m_\chi$ as follows 
\begin{equation}
\label{eq16} 
\epsilon^4=\epsilon_{u.b.}^4\, \exp\left(-\dfrac{L}{X_\chi(\epsilon_{u.b.})}\right)\,, 
\end{equation}
where $\epsilon$ 
is the charge  value of the MCPs  that would reach the detector without energy loss  (so that $s_{up}\propto \epsilon^4$), and 
$\epsilon_{u.b.}$ is the maximum charge value of the MCP that reach the detector with the proper accounting  of their energy losses.

\begin{figure*}[!htb]
    \centerline{
    \includegraphics[width=0.5\textwidth]{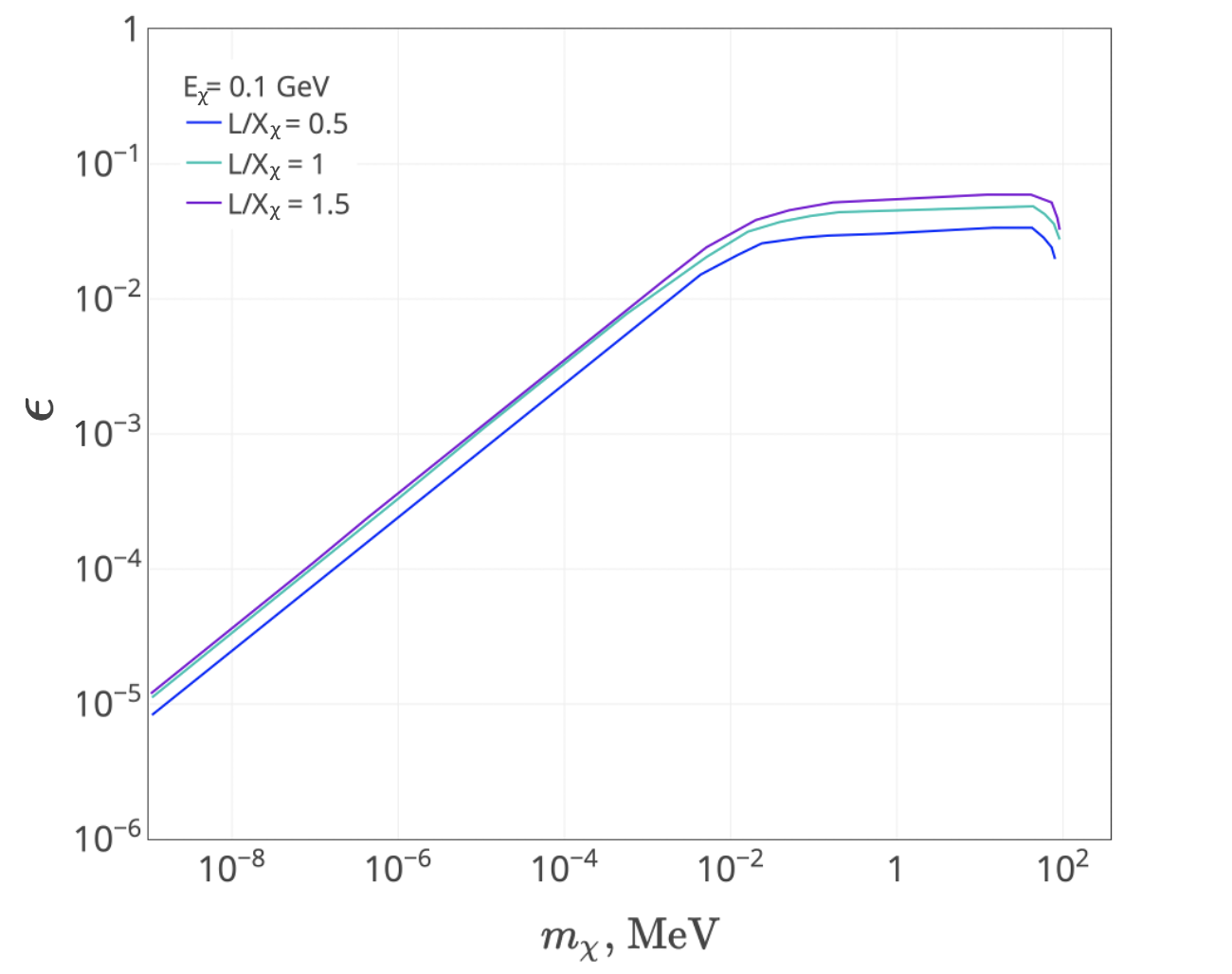}
    \includegraphics[width=0.5\textwidth]{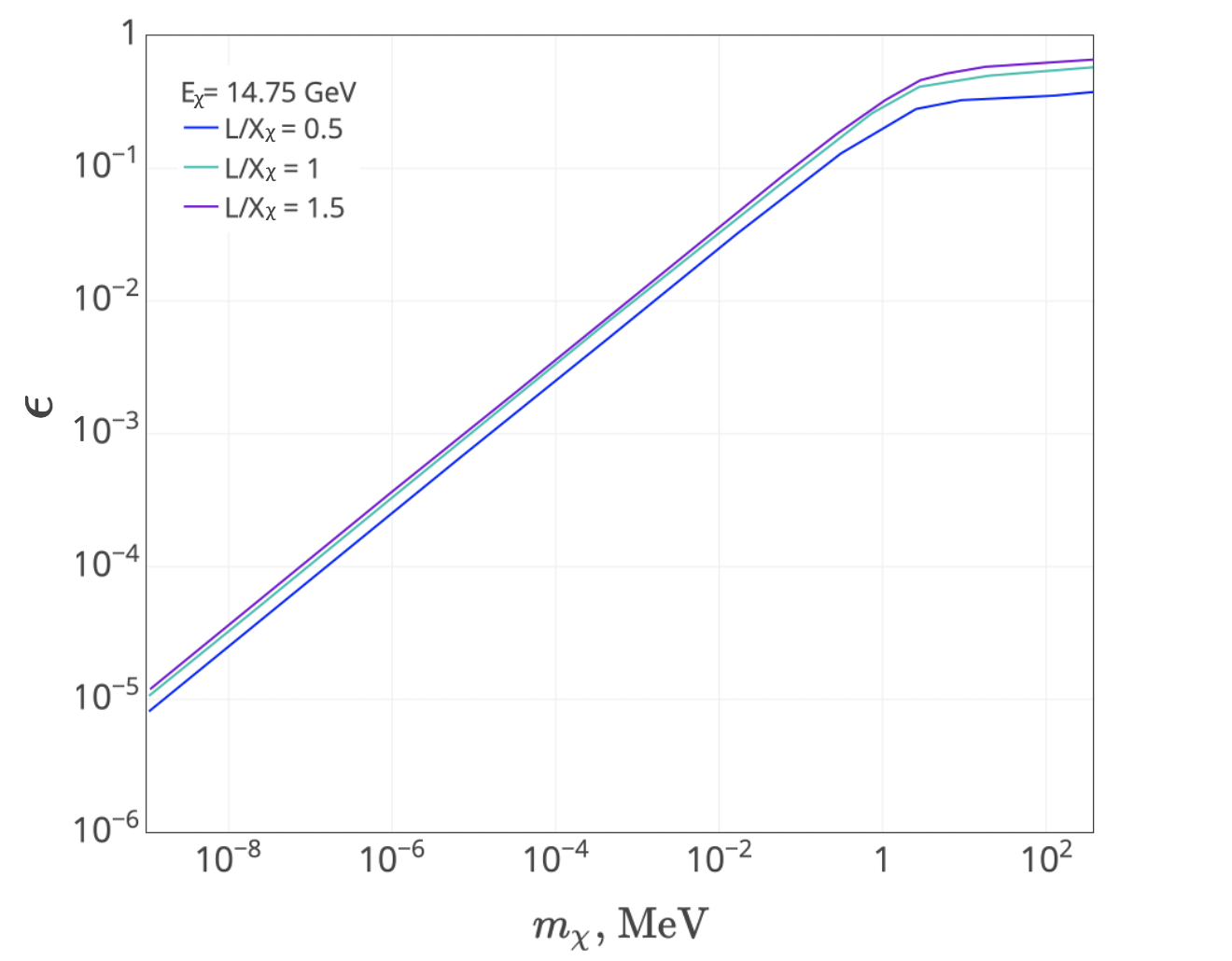}
    }
    \caption{Illustration of the stopping power for two examples of MCP initial energies.
    \label{fig:stopping-power}}
\end{figure*}

 We note that for the relatively large charge, $\epsilon\gtrsim 10^{-2}-10^{-1}$, 
and  $m_\chi \gtrsim m_e$ the rate of MCP production is fairly high. On the other hand, the rate of MCP 
absorption in the dump due to ionization and $e^+e^-$ pair production is also high.  Thus,  the number of produced MCP is compensated  by their attenuation in the dump. 
Moreover, numerical calculations reveal that for $\epsilon \ll 1$ and $m_\chi \lesssim m_e$ 
the radiation stopping power of MCP is not negligible and plays the dominant
role in their absorption in the dump.  In Fig.~\ref{fig:stopping-power} 
the typical 
stopping power and its impact on the limit is illustrated for $X_\chi\simeq L$ and two values of MCP energy at production, which are relevant for SLACmQ experiment. 

%%%%%%%%%%%%%%%%%%%%%%%%%%%%%%%%%%%%
\subsection{Angular and timing acceptance}
%{ \bf Do we need this sentence here?: The angular acceptance of the SLACmQ detector is 2 mrad. } 
There are also two important benchmark conditions that should be taken into account  to constrain 
the charge-mass parameter space of MCP in the case of SLACmQ. 
 The first one is associated with the time interval of data taking. The arrival time of the MCPs in the detector is defined 
%(the main timing window is 250 ns). 
by the muon beam scintillating counters located  between the target and the detector at 
the distance of $S_\mu=82.6$ m from the target ($L_\mu=61$~m through
sandstone) as shown in Fig.\,\ref{setup SLAC}.  The  data are collected during the time interval  $\Delta t_{coll}=250$\,ns. 
The offset of synchronized signal is such that the muon signals came to muon counters in  $\Delta t_{off}=60$~ns after the start of main timing window. It is assumed that the muons are (ultra)relativistic and collisionless, hence they reach the muon counter without any delay with respect to the light. Indeed, 
the typical radiation length of the muon can be estimated as 
$X_0^\mu \simeq X_0^e (m_\mu/m_e)^2 \simeq 4.4\cdot 10^4 X_0^e$,  for 
$X_0^e\simeq \mathcal{O}(1)\,\mbox{cm}$ that implies $X_0^\mu \simeq 440\, \mbox{m}$, so that muon loses 
a small amount of the initial energy when it reaches the detector at the typical distance of $110\,\mbox{m}$. 
On the other hand, the produced in the target MCP of energy $E_i$ can scatter off the sand and reach the muon counter with time delay of 
\begin{equation}
\label{eq20}
   \Delta\tau_\mu=\int\limits^0_{S_{\mu}} dx\dfrac{1-\beta}{\beta\, }=\int\limits_{E_{i}}^{E_{\mu}} \dfrac{dE}{\left|dE/dx\right|}\dfrac{1-\beta}{\beta\, }\,,
\end{equation}
where $E_\mu$ is the MCP energy when it enters the muon counter, and the MCP velocity is $\beta\equiv\sqrt{1-m_\chi^2/E^2}$.  
Then, the MCP has a time slot of 
\[
\Delta\tau_\chi\equiv\Delta t_{coll}-\Delta t_{off}-\Delta \tau_\mu 
\]
to reach the detector while the time window for the data collection is still open. Therefore the MCP must cover the distance $L-S$=$L_\mu-S_\mu=27.5$\,m in a shorter time. This travelling through the sandstone is also accompanied  with additional time delay due to scattering. This energy loss makes MCP less relativistic. Taking this into account we obtain the following condition for MCP to arrive in time and be recorded, 
\begin{equation}
\label{eq20}
   \int\limits_{L_{\mu}}^{L} \dfrac{dx}{\beta\, c}=\int\limits_{E_L}^{E_{\mu}} \dfrac{dE}{\left|dE/dx\right|}\dfrac{1}{\beta\, } < \Delta\tau_\chi\,,  
\end{equation}
where $E_L$ is the MCP energy in the detector.

If the energy losses are small, the energies of MCP in front of the scintillating counters are:
\noindent$E_L\approx E - \left|\dfrac{dE}{dx}\right|\, L, \quad E_\mu \approx E - \left|\dfrac{dE}{dx}\right|\, L_{\mu}$, if not, it is necessary to solve the equation:
\[\int\limits_{E_L}^{E_{\text{i}}} \dfrac{dE}{\left|dE/dx\right|} = L\,,\]
where $dE/dx$ is from Eq.(\ref{eq14}) and takes into account the energy loss in matter. To solve this equation we adopt the approximate expressions from  Appendix\,\ref{appendixB}. 

The second benchmark condition is associated with  the detector angular acceptance of 2\,mr. 
%Note that energy  losses and the multiple scattering is beyond the scope of the present paper.
We estimate  the multiple scattering angle of MCP as a function of energy   it has in front of the detector ($E_L$) by using Eq.~(\ref{eq11}, \ref{eq13}) and require it to be:
\begin{equation}
\label{eq21}
\theta_\chi= \sqrt{\overline{\theta^2_Z}+\overline{\theta^2_e}} < 2 \mbox{ mr}.
\end{equation}

%%%%%%%%%%%%%%%%%%%%%%%%%%%%%%%%%%%%%%%%%%%%%%%%%%%%%%%%%%%%%%
\subsection{Revised exclusion plot}

\begin{figure}[!tbh]
\centering
\includegraphics[width=0.5\textwidth]{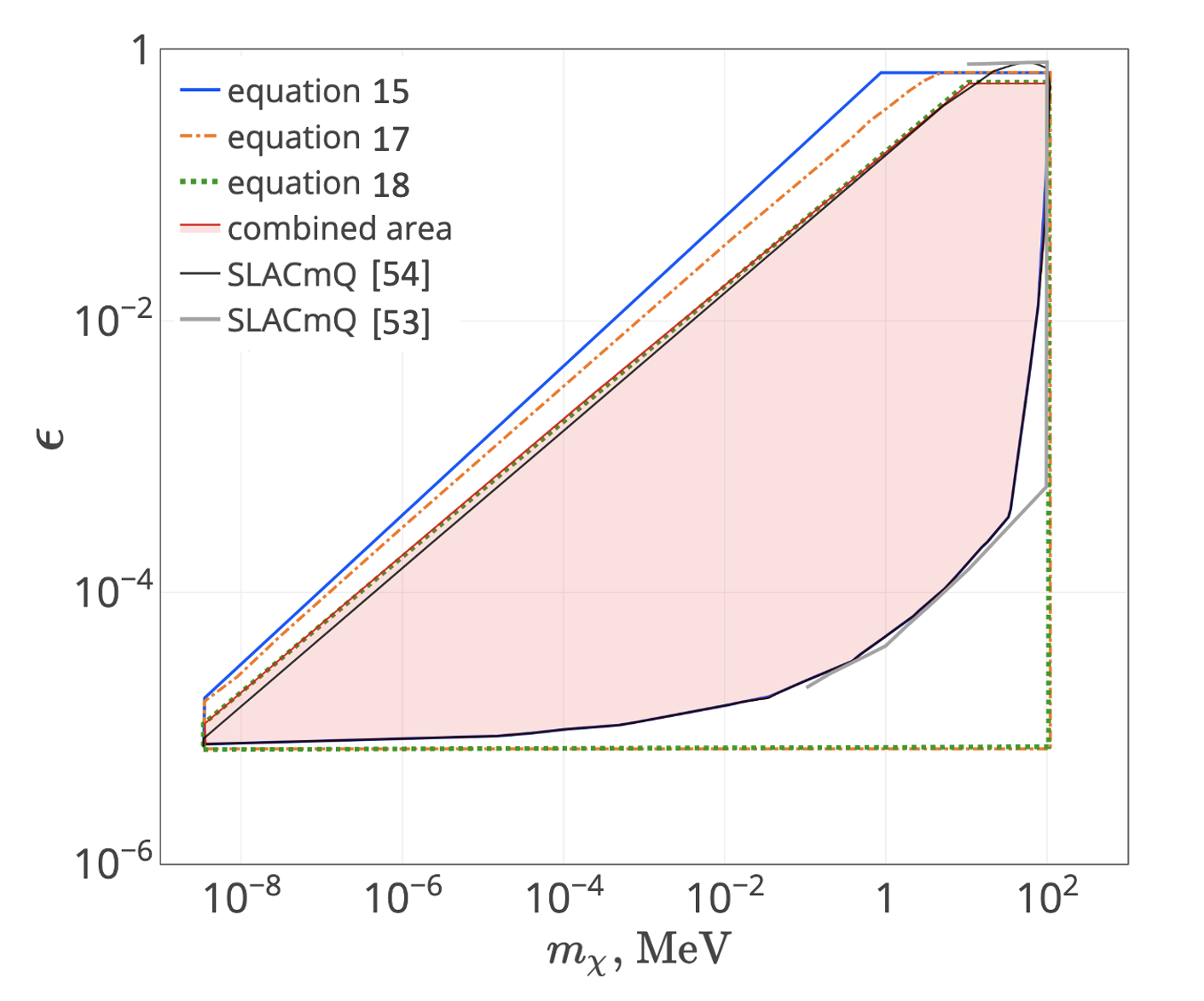}
\caption{Regions bounded from Eqs. (\ref{eq16}, \ref{eq20}, \ref{eq21}) and the combined exclusion region from our analysis of the SLACmQ data. The obtained region closely matches the result of Ref.\,\cite{Prinz:2001qz} ({\it solid black line}) and refines that published in the original SLACmQ paper\,\cite{Prinz:1998ua} ({\it solid grey line}). 
\label{Combinedexclusion} }
\end{figure}

We have explained previously,  all the steps towards reevaluation of the SLACmQ bounds on models with MCP. All the numerical calculation we perform with the approximate spectrum of MCP presented in Ref.\,\cite{Prinz:2001qz}.  
In Fig.~\ref{Combinedexclusion} 
we show the resulted exclusion region in ($m_\chi, \epsilon_\chi$) plane which is 
determined by Eqs.~\eqref{eq16}, \eqref{eq20}, \eqref{eq21}.  Expectedly, the MCP scatterings are mostly important for relatively large charges and small masses. Most critical are deflections of the MCP trajectories depicted by the green dashed line. 

The obtained in this way region is similar to the one presented in Ref.\,\cite{Prinz:2001qz}. As one can see from Fig.\,\ref{Combinedexclusion}, the exclusion region from the original publication\,\cite{Prinz:1998ua} of the SLACmQ collaboration is rather limited in mass and has no upper bounds. What is more important, there is a triangular region at $\epsilon\sim 10^{-3}$ and $m_\chi\sim100$\,MeV, which we find not constrained from the SLACmQ data, contrary to their publication. This region is also open is Ref.\,\cite{Prinz:2001qz}.   It is worth mentioning again that we adopt the lower edge of the  curve 
SLACmQ~\cite{Prinz:2001qz} shown in Fig.~\ref{Combinedexclusion} from Fig.~5.17 of the 
Ref.~\cite{Prinz:2001qz}. The numerical   values of the regarding coupling correspond to the $\epsilon$ in Eq.~(\ref{eq16}). 

%1) solution of the equation (Figure \ref{Combinedexclusion}):
%\begin{equation}\label{eq19}
%\epsilon_\chi^4\, \exp\left(-\dfrac{L}{X_\chi(\epsilon_\chi)}\right) = \epsilon^4\, ; \quad %X_\chi=\int^{E_{\text{max}}}_{E_{\text{min}}} %\int^E_{E_{\text{cut}}}\dfrac{dE'}{\left|\dfrac{dE}{dx}(E')\right|}\, \dfrac{dN}{dE}(E)\, dE
%\end{equation}

%2) the time interval (eq.\ref{eq17}) (Figure \ref{Combinedexclusion}):
%\begin{equation}\label{eq20}
%\tau_\chi=\int^{E_L}_{E_\mu}\dfrac{dE}{dE/dx}\,\dfrac{1}{\beta\, c}dE < 190 ns
%\end{equation}

%3) the deflection angle (eq.\ref{eq18}) (Figure \ref{Combinedexclusion}):
%\begin{equation}
%\theta_\chi = \sqrt{\overline{\theta^2_Z}+\overline{\theta^2_e}}< 2 \mbox{  mr}
%\end{equation}

%{\bf  Do we need this part below for  the calculation?: }

%...ionization energy $I=8.15$ eV. 
% The 
%detector consists of 4 scintillating counters sensitive to signals as small as a single excitation or ionization with  %threshold about $1/3$ photoelectron signal. 
%The ratio of the number of photoelectrons to deposited in the 
%scintillation energy derived is 0.317 $\pm$ 0.026 photoelectrons/keV \cite{Prinz:1998ua}. The deposited energy in
%scintillator could be calculated from solving equation (\ref{eq3}):
%$\left(-\dfrac{dE}{dx}_{\text{ion}}\right)_{E=E_{\text{cut}}}=1.05 \text{ keV}$.
%The length for MCP ionization within detector is 1.31 m.

%%%%%%%%%%%%%%%%%%%%%%%%%%%%%%%%%%%%%%%%%%%%%%%%%%%%%%%%%%%%%%%%%%%%%%%%%%%%%%%%%

\section{Calculation of the double-differential cross section of MCP pair production
\label{NA64Cross-section}}

\begin{figure}[!htb]
\centering
\label{fig:my_label-4}
\includegraphics[width=0.5\textwidth]{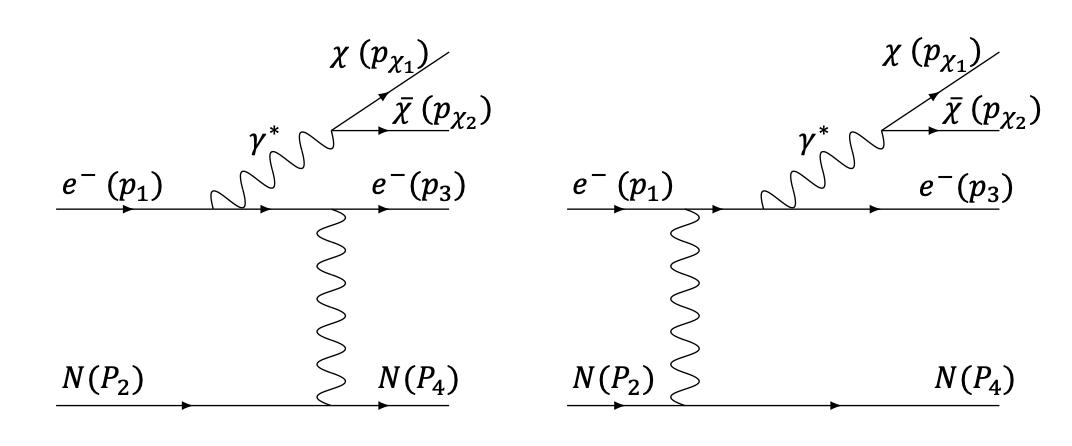}
\caption{Leading Feynman diagrams for MCP pair production process.  
\label{MCP production} }
\end{figure}

In this Section we describe our procedure of calculation the double-differential cross section of MCP production 
in electron scattering off nucleus. At the sufficiently small MCP charge the production happens mostly through 
emission of the virtual photon,  
\begin{align}
    \label{process}
& e(p_1) N(\mathcal{P}_2) \to e(p_3)
\gamma^*(p_{\chi_1}+p_{\chi_2}) N(\mathcal{P}_4) \to \\
& \to e(p_3) \chi(p_{\chi_1}) \bar{\chi}(p_{\chi_2}) N(\mathcal{P}_4)\,, \nn 
\end{align}
%\begin{equation}
%\label{process}
%e(p_1) N(\mathcal{P}_2) \to e(p_3)
%\gamma^*(p_{\chi_1}+p_{\chi_2}) N(\mathcal{P}_4) \to e(p_3) \chi(p_{\chi_1}) \bar{\chi}(p_{\chi_2}) %N(\mathcal{P}_4)\,,
%\end{equation}
and since it is purely electromagnetic process we calculate it in the exact tree-level approach. The Feynman diagrams referring to the tree-level amplitude are presented in 
Fig.\,\ref{MCP production}.  
This contribution is linear in $\epsilon$, and we neglect contributions proportional to $\epsilon^2$ coming from other diagrams. 

The NA64$e$ experiment utilizes the electron beam of 
$E_1=100$~GeV and lead target (see below Sec. \ref{NA64SensitivitySection}).
 We carry out the integration of squared exact tree-level amplitude 
 over the phase space  
of the process \eqref{process}  by exploiting the latest version of the {\tt CalcHEP} package~\cite{Belyaev:2012qa}.  
The vertex for $NN\gamma$ interaction we take to be  
 \begin{equation}
 \label{FF1}
i e Z F(-q^2) \gamma_\mu\,.
 \end{equation}
It corresponds to the spin-$1/2$ nucleus interaction with photon, 
here $q=(\mathcal{P}_2-\mathcal{P}_4)$ is nucleus transfer momentum, 
$\mathcal{P}_2$ and $\mathcal{P}_4$ are initial and final momenta of the nucleus.  
The elastic  form-factor  $F(-q^2)$ has the following form 
\begin{equation}
\label{FF2}
F(t)=\frac{a^2 t}{(1+a^2 t)} \frac{1}{(1+t/d)}  \,,  
\end{equation}
where $t=-q^2$ is the squared transfer momentum, $a=111 Z^{-1/3}/m_e$ and 
$d=0.164 A^{-2/3}\, \mbox{GeV}^2$ are screening and nucleus parameters 
respectively. For the lead active target (atomic number $A=207$, nuclear charge $Z=82$) of NA64$e$ one estimates the following typical momenta transfer associated with 
screening effects and nucleus size respectively: 
$\sqrt{t_{a}}=1/a\simeq 2\cdot 10^{-5}$~GeV and 
$\sqrt{t_d}=\sqrt{d}\simeq 6.7\cdot 10^{-2}$~GeV.

\begin{figure}[!tbh]
\centering
\includegraphics[width=0.45\textwidth]{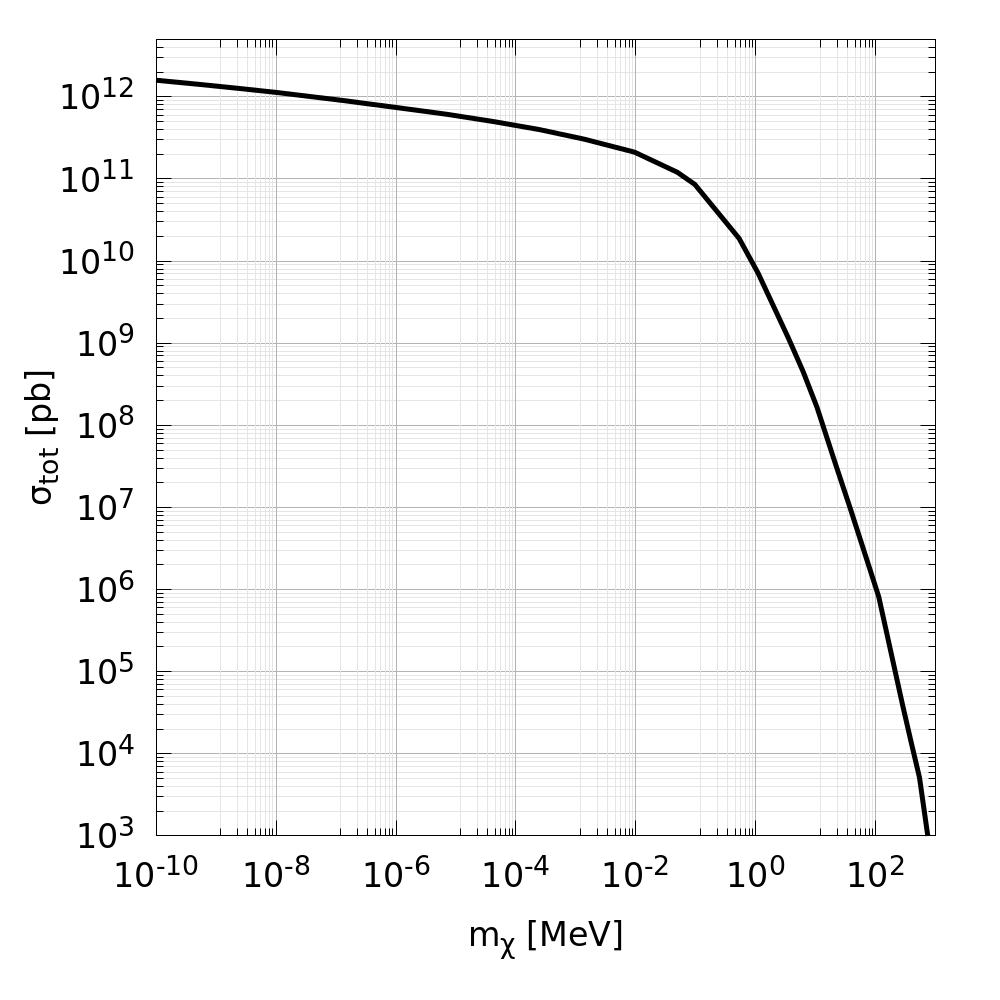}
\caption{Total cross section of MCP pair production as the function of $m_{\chi}$ for $E_{beam}=100$~GeV and $\epsilon=1$. 
\label{NA64TotCS} }
\end{figure}

We implement the form factor \eqref{FF1},\,\eqref{FF2} in the C++ code of
{\tt CalcHEP} and carry out {\tt VEGAS} Monte-Carlo integration of the cross section for various masses $m_\chi$ in order to 
obtain the differential  energy spectra of the produced MCP. Before presenting its form let us note, that to check our calculations we 
integrate the double-differential cross section over the MCP energies and thus evaluate the total MCP production cross section which is shown in  Fig.~\ref{NA64TotCS} as a function of MCP mass  $m_\chi$. 
Similar plot can be found in Fig.\,3 of Ref.\,\cite{Chu:2018qrm}, where the calculation is done
in the limit of massless electron and for $m_\chi>1$\,MeV. We have checked that in this limit
our estimate of the total cross section matches with that of Ref.\,\cite{Chu:2018qrm} with 
accuracy  of a few percent.\footnote{We thank J.\,Pradler, X.\,Chu and A.\, Pukhov for helping 
us to do this cross-check.}  Therefore, in our study we extend the previous work on (much) lighter 
MCP and account corrections due to non-zero electron mass.    

%%%%%%%%%%%%%%%%%%%%%%%%%%%%%%%%%%%%%%%%%%%%%%%%%%%%%%%%%%%%%%%%%%%%%%%%%%%%%%%%%%
\section{Sensitivity of NA64$e$ 
\label{NA64SensitivitySection}}
\begin{figure*}[tbh!]
\includegraphics[width=0.9\textwidth]{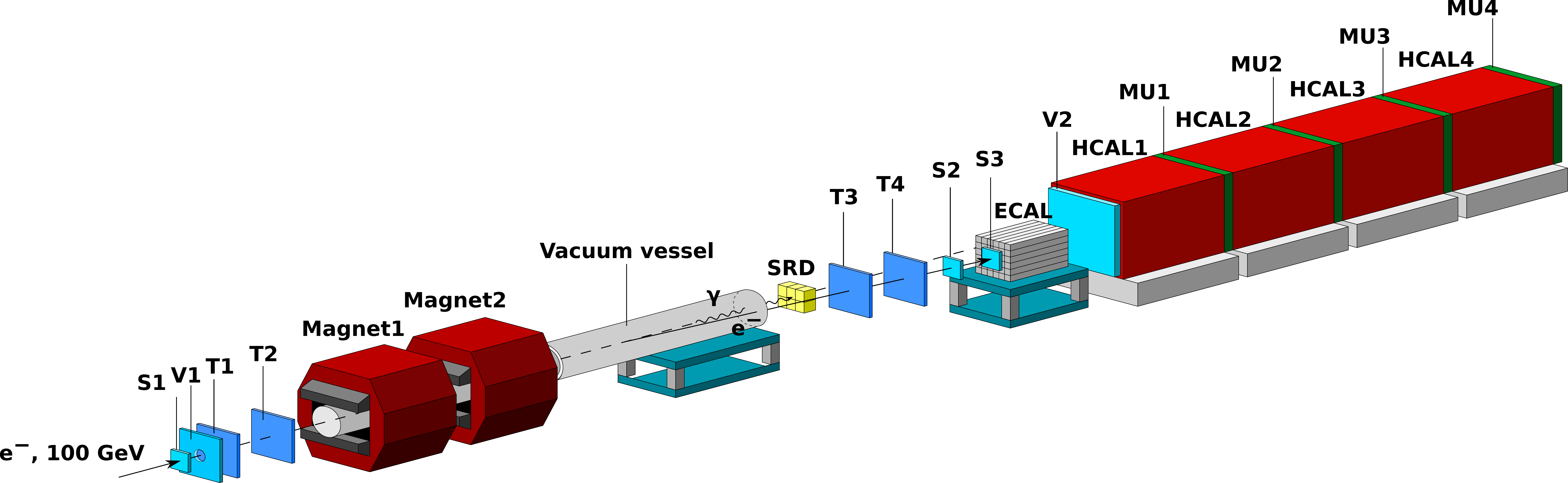}%
\caption{Schematic illustration of the setup to search for MCP production in the reaction  $e^-N\rightarrow e^-N \gamma^* (\to \bar{\chi} \chi)$ of 100 GeV e$^-$
 incident  on the active ECAL target. See text.}
 \label{setup}
\end{figure*} 
%%%%%%%%%%%%%%%%%%%%%%%%%%%%%%%%%%%
\par  Before evaluating the MCP search sensitivity with the  NA64e experiment, let us briefly describe its main features relevant for further discussion.
  The  NA64e detector  is schematically shown in Fig.~\ref{setup}.
The experiment  employed  the optimized H4  100 GeV  electron beam from the CERN SPS.  
The  beam  has a maximal intensity $\simeq 10^7$ electrons per  SPS spill of 4.8 s produced by the primary 400 GeV proton beam  with an intensity of few 10$^{12}$ protons on target.  The detector utilized 
the beam defining  scintillator (Sc)  counters $S_{1-4}$ and veto  $V_{1,2}$, a magnetic spectrometer  consisting of two successive  dipole Magnets$_{1,2}$ with the integral  magnetic field of $\simeq$7 T$\cdot$m  and a low-material-budget tracker. The tracker was a set of two upstream Micromegas chambers T$_{1,2}$,  and two downstream MM$_{3,4}$,   allowing the measurements of $e^-$ momenta with the precision $\delta p/p \simeq 1\%$. 
To significantly improve the electron identification,  the synchrotron radiation (SR) emitted in the magnetic field of the Magnets$_{1,2}$ 
 was used for their  tagging with a SR detector (SRD) \cite{Gninenko:2013rka, Depero:2017mrr}. By using this method the initial  fraction  of the hadrons in the  beam $\pi/e^- \lesssim 10^{-2}$ was further suppressed by a factor $\simeq 10^3$.   
The detector was also equipped with an active target, which is an  electromagnetic calorimeter (ECAL),  a  matrix of $6\times 6 $  Shashlik-type counters  assembled from  lead and scintillator plates  for  measurement of the electron energy  $E_{ECAL} $. 
 Each counter has $\simeq 40$ radiation  lengths ($X_0$) with the first 4$X_0$ serving as a preshower detector.   
 Downstream of the ECAL, the detector was equipped with a large  high-efficiency veto counter 
 VETO, and a massive, hermetic hadronic calorimeter (HCAL) of $\simeq 30$ nuclear interaction 
 lengths in total. The modules HCAL$_{1-3}$  provided  an efficient veto to detect muons or 
 hadronic secondaries produced in the $e^- A$ interactions  in the target.   The search 
 described in this paper uses the data sample of $N_{EOT}=2.84\times 10^{11}$ EOT collected  in 
 the three  years 2016, 2017 and 2018.  
 The method, briefly discussed in Sec. I and  proposed in Refs.~\cite{Gninenko:2013rka}, 
 is based on the detection of the missing energy, carried away by the hard bremsstrahlung 
  MCPs produced in the process $e^- N \to e^- N \gamma^*( \to\bar{\chi} \chi) $  of 
  high-energy electrons scattering in the active ECAL  target.   The advantage of NA64 idea compared 
  to the beam dump one is that its sensitivity is proportional to $\epsilon^2$. The latter is associated 
  with the $\chi$ production  and its subsequent prompt escaping the 
 detector without interactions in the HCAL modules.

 In the following, similar to  Ref.~\cite{Chu:2018qrm}, we first  calculate the number 
of missing energy  events associated with MCP emission by energetic electrons incident on the thin target. The latter  
implies a single scattering on average and then a rapid degrade of the electron energy due to bremsstrahlung. For the case of NA64$e$,  we assume that the incident electron 
produces  MCP in the first scattering within the first radiation length of the ECAL  
target of NA64$e$, $ X_0 \simeq 0.56$\,cm. This assumption is justified by the fact that the high energy electron loses most its energy in a single process transferring it to the emitted photon.     

Now let us 
evaluate the sensitivity of NA64$e$ to model parameters of MCP taking into account the energy loss of the 
millicharged particles in the detector. This energy transfers to the electromagnetic channel which NA64$e$ closely monitors and hence sums with all other sources of electromagnetic activity. The experimental signature of the MCP detection, would be an event with  the missing energy $E_{miss} \gtrsim 50$ GeV, see, e.g. Ref.\cite{Banerjee:2019pds}. The events with lower missing energy are not considered as potentially signal events. Therefore, the estimate of MCP energy loss within the detector is an important step towards understanding  prospects of NA64$e$ in testing models with MCP.  

\begin{figure*}[!htb]
\centering
\includegraphics[width=0.45\textwidth]{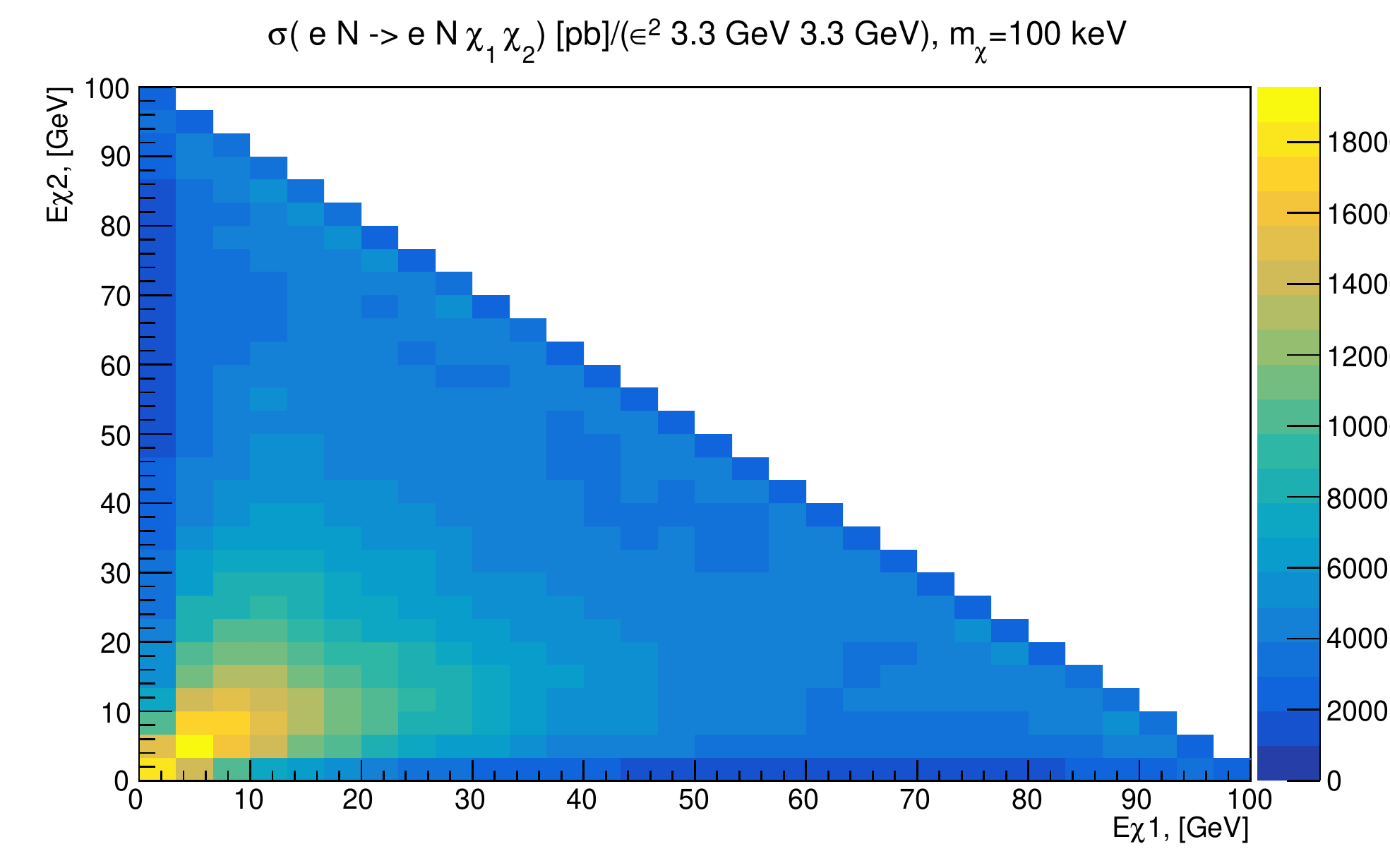}
\includegraphics[width=0.45\textwidth]{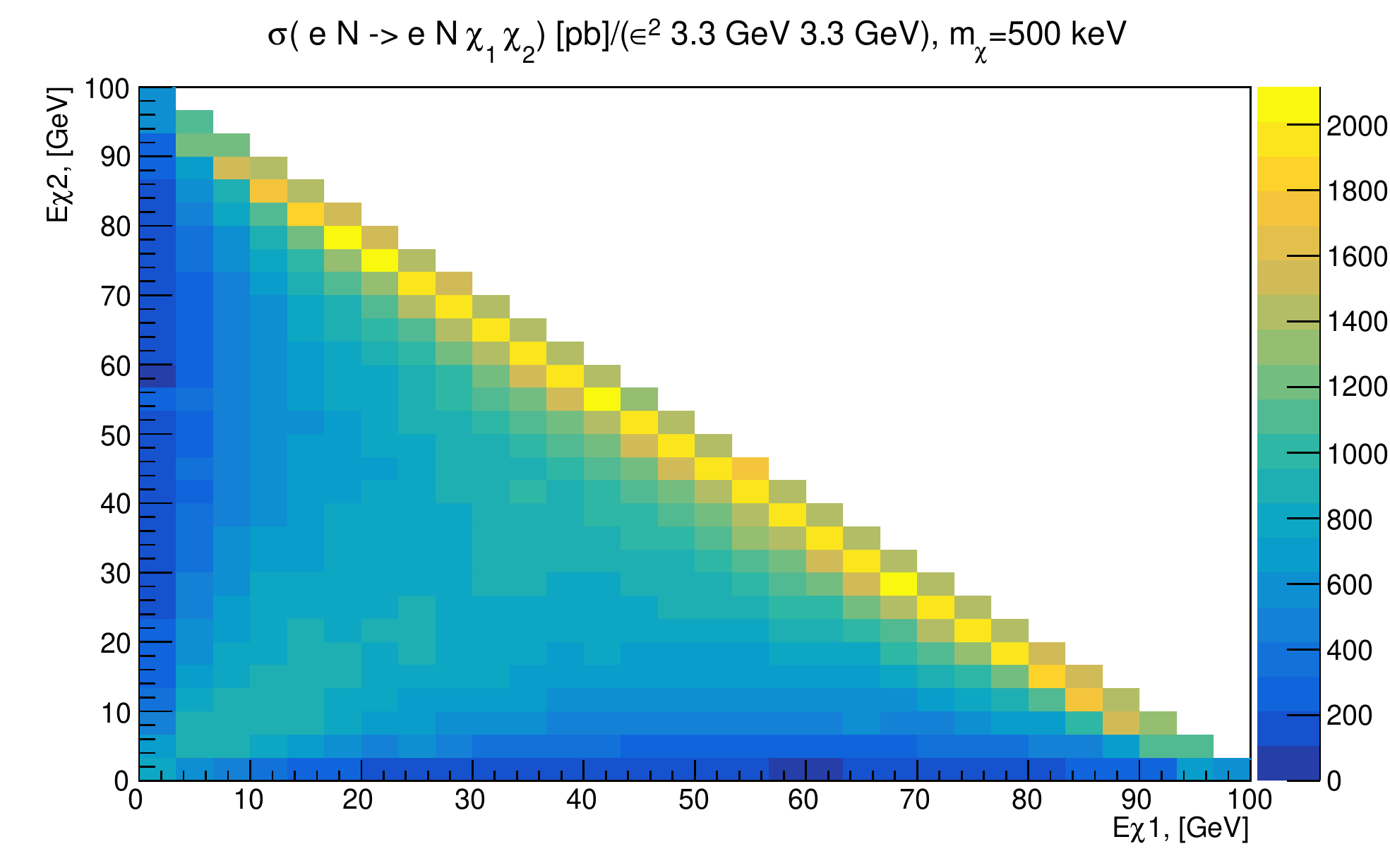}
\includegraphics[width=0.45\textwidth]{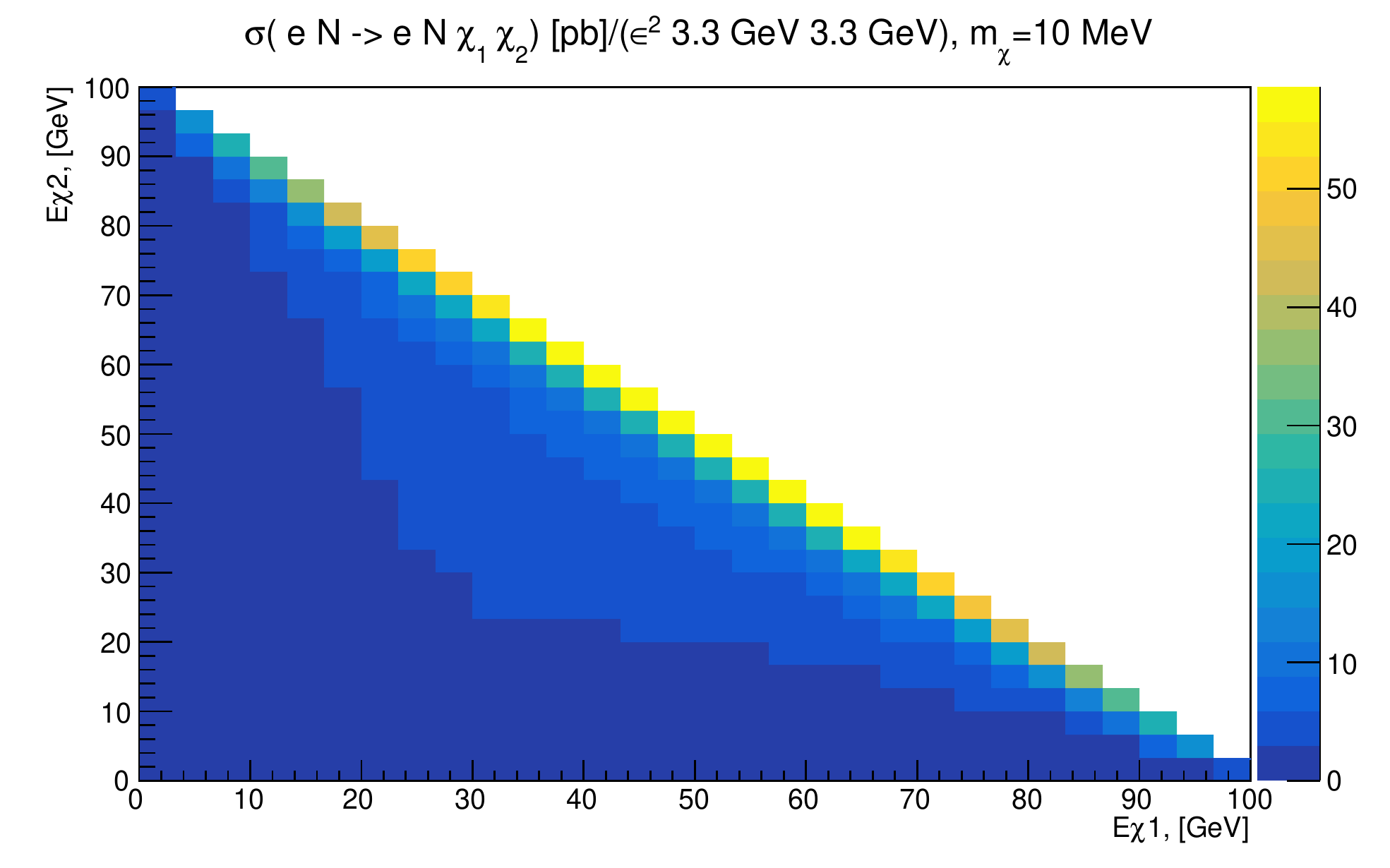}
\includegraphics[width=0.45\textwidth]{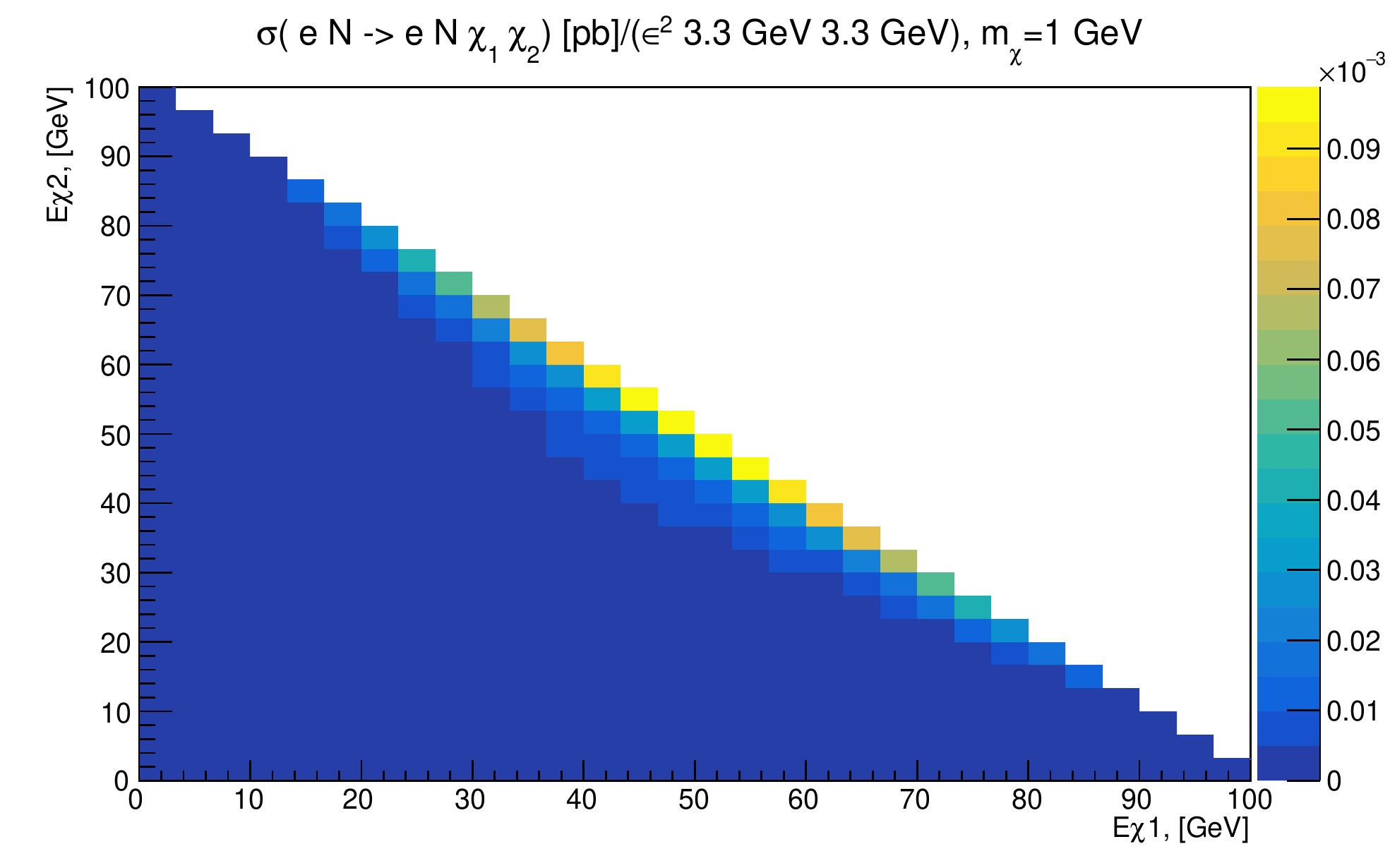}
\caption{Double-differential cross section of the MCP pair production
as a function of MCP energies $E_{\chi_1}$ and $E_{\chi_2}$ for a set of MCP masses $m_\chi$ and energy of incident electron $E_{beam}=100$\,GeV. Note that only events above the line $E_{\chi_1}+E_{\chi_2}>50$\,GeV are interesting with cuts adopted in NA64$e$ experiment. 
\label{2DSpectra}
}
\end{figure*}

The total number of MCP pairs produced in the 
process $eN\to eN \chi\bar{\chi}$ can be calculated as follows 
\begin{equation}
N_{\chi\bar{\chi}}  = \frac{\rho N_A }{M_A} N_{EOT} X_0
\int\limits_{\mbox{s.b.}} dE_{\chi_1} dE_{\chi_2}
\frac{d \sigma (E_{beam}) }{d E_{\chi_1} d E_{\chi_2}}\,. 
\label{NsigEvThickTarg}
\end{equation}
%\begin{equation}
%N_{\chi\chi}  = \frac{\rho N_A }{A} N_{EOT} 
%\int\limits_{E^{th}_{miss}}^{E_{beam}} d E_{es}  \frac{d l_{e s}}{d E_{es}}  (E_{es})
%\int\limits_{\mbox{s.b.}} dE_{\chi_1} dE_{\chi_2}
%\frac{d \sigma (E_{es}) }{d E_{\chi_1} d E_{\chi_2}} 
%\label{NsigEvThickTarg}
%\end{equation}
Here  $N_{EOT}$ is a number of  electrons accumulated on
target,
$\rho = 11.34\,~\mbox{g}\, \mbox{cm}^{-3}$ is a density of the lead ECAL active 
target,  $M_A=207~\mbox{g}\, \mbox{mol}^{-1}$ is the atomic mass of the target, $N_A$ is Avogadro's number,  $E_{beam}=100$\,GeV is the energy of initial electron from the beam,
the double-differential cross section
of $\chi\bar{\chi}$ production $d \sigma /(dE_{\chi_1} dE_{\chi_2})$ is calculated 
by using {\tt CalcHEP} as described in Sec.\,\ref{NA64Cross-section}. The diplots corresponding to several choices of MCP mass are presented in Fig.\,\ref{2DSpectra}.
Naturally, the distribution is symmetric with respect to interchange of the positive and negative MCP.

\begin{figure}
    \centering
    \includegraphics[width=0.45\textwidth]{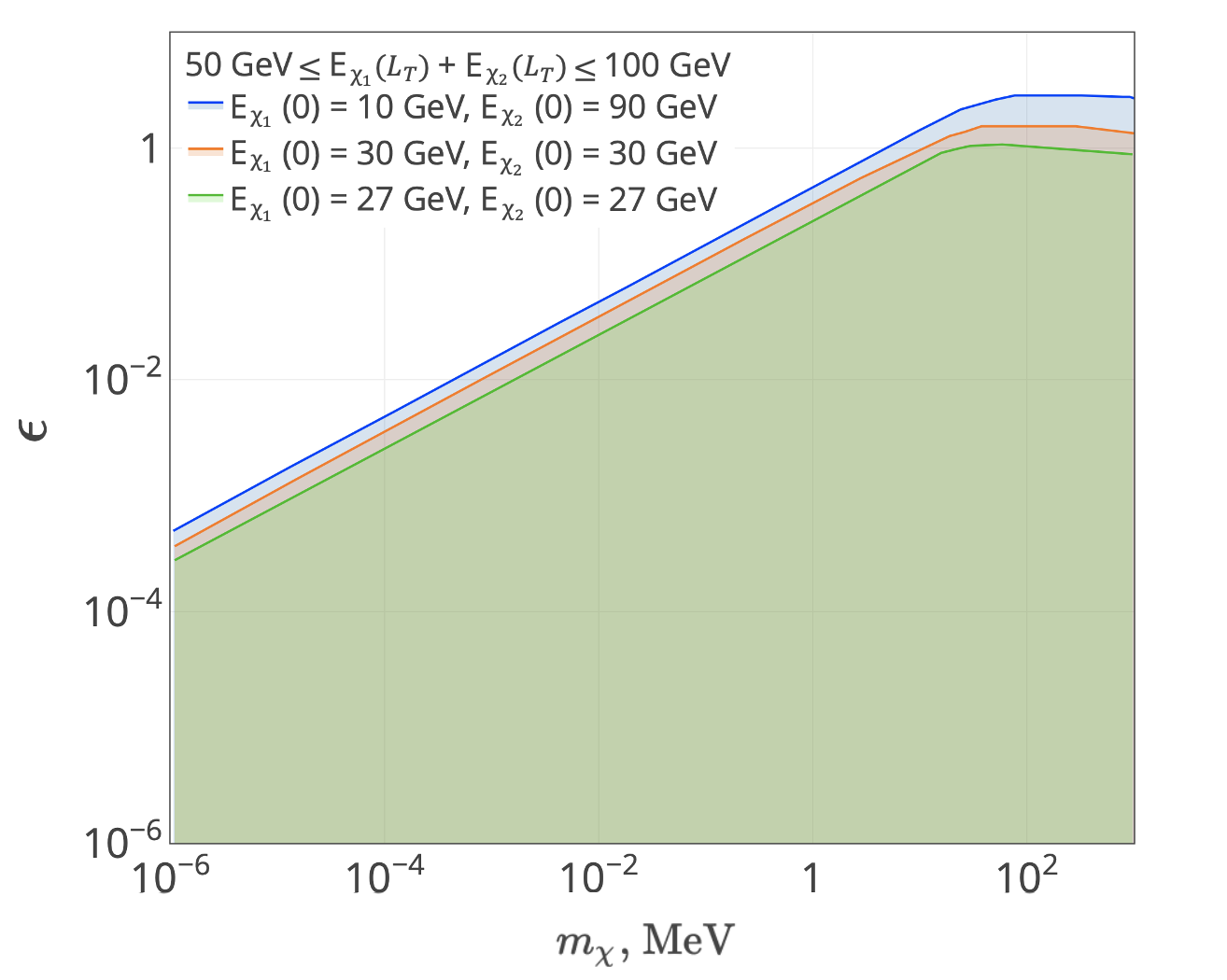}
    \caption{The MCP pairs with a chosen initial energies do not contribute to the signal events for parameters above the corresponding lines.}
    \label{fig:Loss-NA64}
\end{figure}

The integration over $E_{\chi_1}$ and $E_{\chi_2}$ in \eqref{NsigEvThickTarg} is 
performed inside the signal box (s.~b.). This region is determined by  the
 missing energy cut of the electromagnetic calorimeter (ECAL) exploited in NA64$e$, $E_{miss}^{th}\equiv 50$~GeV, and the initial energy of the electron 
 beam  $E_{beam}=100$\,GeV as follows,  
 \begin{equation}
     E_{miss}^{th}  \lesssim  E_{\chi_1} + E_{\chi_2} - \Delta E_{\chi_1} -  \Delta E_{\chi_2} \lesssim E_{beam}\,.
 \end{equation}
 Here $E_{\chi_1}$ and $E_{\chi_2}$ are the MCP energies at production,
  $\Delta E_{\chi_1}$ and $\Delta E_{\chi_2}$ are the total energy depositions
 of the millicharged particles in the ECAL due to possible electromagnetic rescattering on their way out. These quantities are  
 estimated through the differential energy loss as follows
    $$
    \Delta E_{\chi_{1,2}} = - \int\limits^{L_{T}}_0 \frac{d E_{\chi_{1,2}}}{dx} dx = E_{\chi_{1,2}}(0)-  E_{\chi_{1,2}}(L_{T}) \geq 0\,,
    $$
and the integral goes over the effective width of the lead material in ECAL, $L_T=22.5$\,cm. The MCP energies at the exit from ECAL $E_{\chi_{1,2}}(L_T)$ can be expressed through the energy of MCP at  
production $E_{\chi_{1,2}}(0)= E_{\chi_{1,2}}$ by  solving the equation 
\begin{equation}
\label{eq:rad-length}
\int\limits^{E_{\chi_{1,2}}(L_T)}_{E_{\chi_{1,2}}(0)}  d E_{\chi_{1,2}} \frac{1}{(dE_{\chi_{1,2}}/dx)} = -L_T,
\end{equation}
where for $(dE_{\chi_{1,2}}/dx)$ we use numerical approximations collected in Appendix\,\ref{appendixA}. 
Finally, for the given  energies of MCP at production in the target,  
$E_{\chi_{1,2}}(0)$,  one has the following region for the signal box 
    \begin{equation}
    E_{miss}^{th} \lesssim E_{\chi_1}( L_{T}) +E_{\chi_2}( L_{T})  \lesssim E_{beam}\,.
    \label{signalBoxDefinition}
    \end{equation}
The relevance of the energy transfer to the electromagnetic cascade is demonstrated in Fig.\,\ref{fig:Loss-NA64}. 
There we chose three examples of MCP pairs with total energy exceeding the threshold value in NA64$e$. In the regions above the corresponding lines the MCP lose its energies emitting energetic photons, which are collected by the NA64$e$ ECAL. Finally the total MCP energy drops below the threshold, and so they do not contribute to the signal events.  
    
Numerical calculations reveal that for the wide ranges of 
$\epsilon \lesssim 0.3$ and $m_\chi \lesssim 1$\,MeV the radiation energy loss of the millicharged 
particle (bremsstrahlung)  dominates over the energy loss due to ionization, $\chi e \to \chi e$, 
and due to pair production, $\chi N \to \chi N e^+ e^-$. 
The energy loss of the millicharged particle due to the 
bremsstrahlung, $\chi N \to \chi N \gamma$, inside the Pb-target can be approximated for the interesting energetic MCP as follows (see Appendix\,\ref{appendixB})  
\begin{align}
     &   \frac{1}{E_\chi} \frac{d E_\chi}{ dx}   \simeq -(X_0)^{-1} \, \epsilon^4    \label{BremsApprox1} \left(\frac{m_\chi}{m_e}\right)^{-2} \\
    &    \simeq - 0.45\, \mbox{cm}^{-1}\, \epsilon^4 \left(\frac{m_\chi}{ 1\,\mbox{MeV}}\right)^{-2}. \nn
\end{align}
It helps to solve \eqref{eq:rad-length} for this region of the model parameter space. 
We note that for $\epsilon=1$ and $m_\chi=m_e$ the result for the electron radiation length $X_0~\simeq~0.56$~cm is restored from 
Eq.~(\ref{BremsApprox1}). It implies the well known result that the electron energy is reduced by factor of 
$e\simeq 2.71$ within the first  radiation  length in lead due to the radiation loss.

\begin{figure}[!t]
\centering
\includegraphics[width=0.5\textwidth]{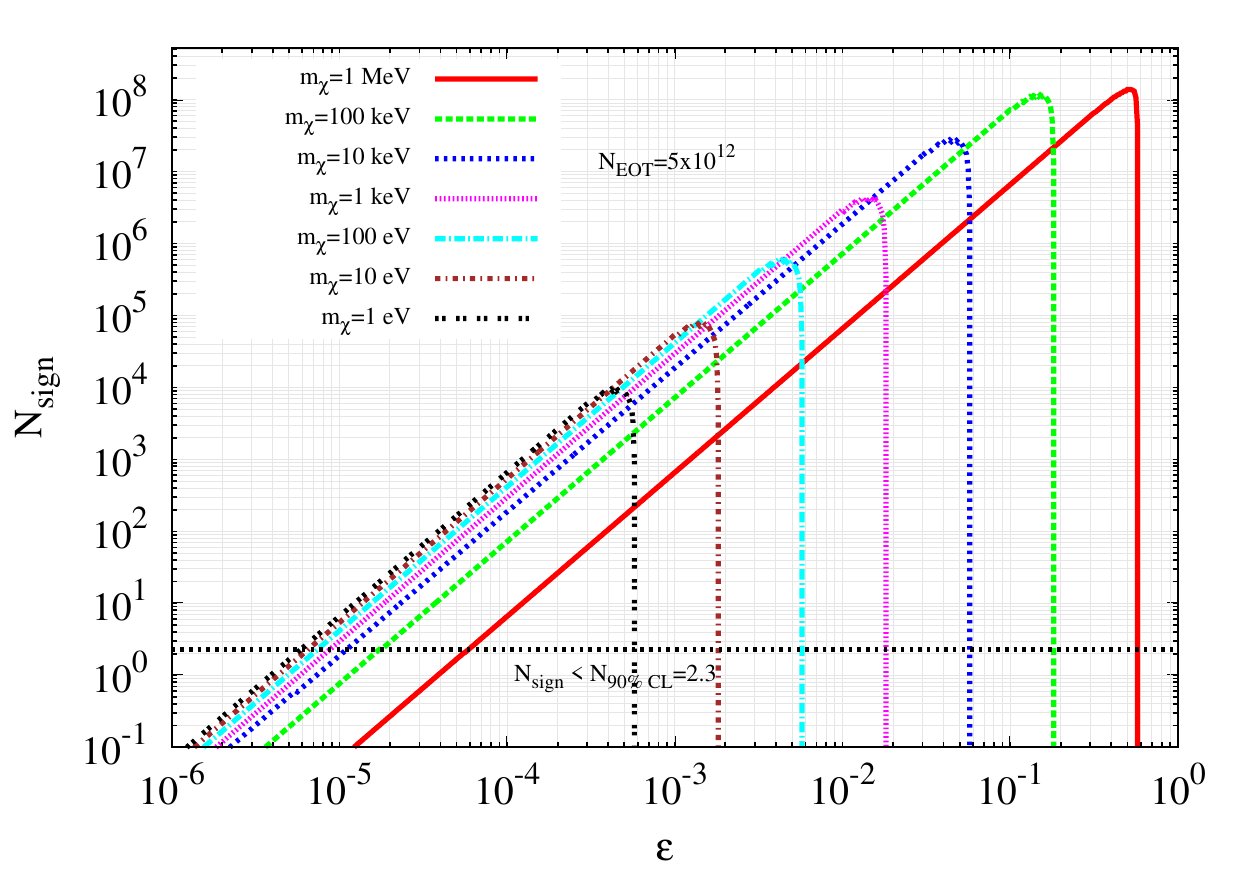}
\caption{Number of signal events $e N \to e N \gamma^*(\to \chi \bar{\chi})$
as function of $\epsilon$. 
\label{NA64ScanOverEpsilon1} }
\end{figure}

Estimated in this way numbers of the signal events are outlined in Fig.\,\ref{NA64ScanOverEpsilon1}
for a set of MCP masses and for the projected statistics of $N_{EOT}=5\times 10^{12}$.

\begin{figure}[!htb]
\centering
\includegraphics[width=0.45\textwidth]{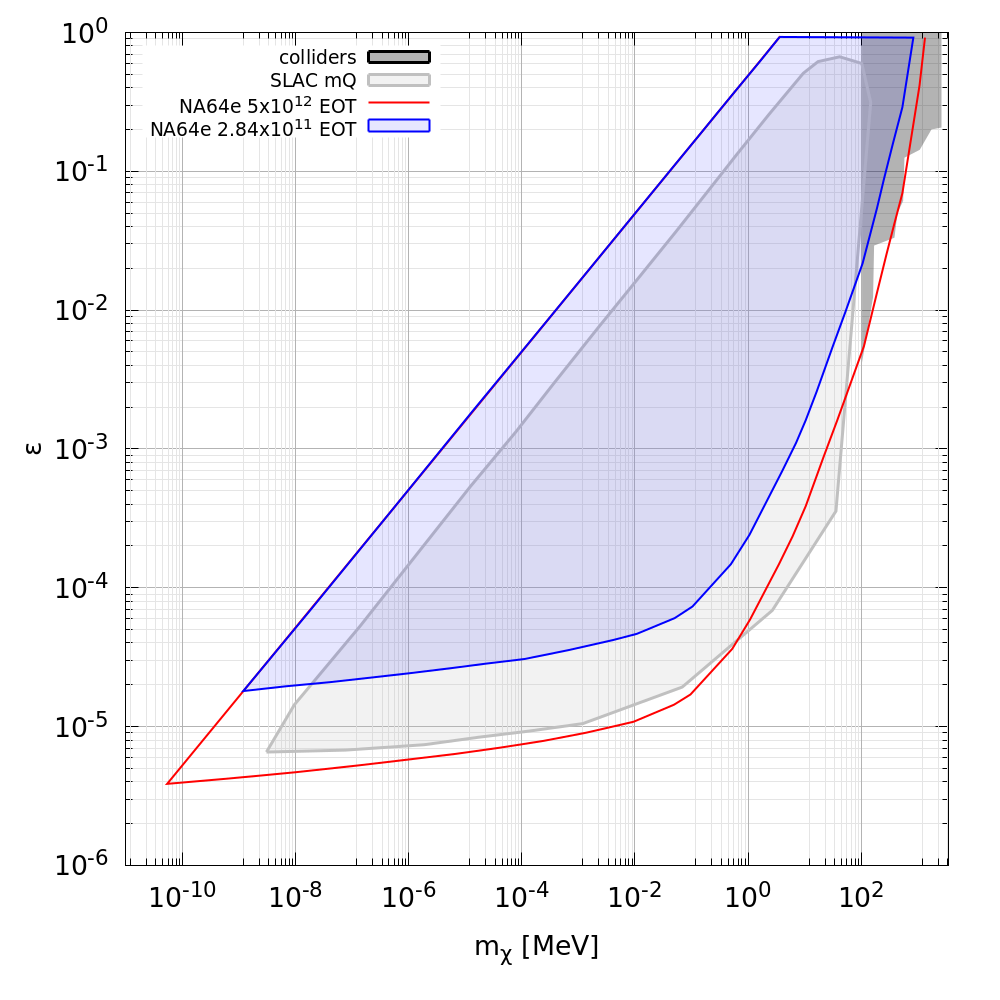}
\caption{Sensitivity contours (at 90\%CL) of NA64$e$ in the $(\epsilon, m_\chi)$ plane for 
 the process $e N \to e N \gamma^*(\to \chi \bar{\chi})$.
 We account for  non-zero electron mass in the calculation, $m_e \neq 0$,
and suppose that MCP pairs are produced
within the first layer (its width is about the electron radiation length) of the lead target. We also set $\alpha_{QED}=1/137$.
\label{NA64Exclusion1} }
\end{figure}

  We evaluate the lower 90\%\,C.L.  bound  on the 
coupling $\epsilon$ by requiring $N_{\chi\bar{\chi}} > s_{up} \equiv 2.3$, which in the Poisson Statistics  
corresponds to the null result at zero background (that is a realistic assumption for NA64$e$ and moderate number of incident electrons we utilize).  The results are shown in Fig.\,\ref{NA64Exclusion1} 
for the 
currently accumulated  $N_{EOT}=2.84\times 10^{11}$ and for the 
projected statistics $N_{EOT}=5\times 10^{12}$ \cite{NA64:2016oww}.

 Concluding this Section we note that it is worth to carry out similar analysis for the MCP 
sensitivity of the 
muon fixed target experiments such as NA64$\mu$ and $M^3$ (see e.~g. 
Refs.~\cite{Sieber:2021fue,Kirpichnikov:2021jev,Kahn:2018cqs,Capdevilla:2021kcf}). In particular, the 
signal of muon missing energy events due to the MCP emission 
can be originated from the following process  $\mu N\to \mu N \chi \bar{\chi}$, for which the effects of 
the MCP passage  through the  detector should be taken into account.  
This task is beyond the scope of the present paper  and we leave it  for  future study. 

%%%%%%%%%%%%%%%%%%%%%%%%%%%%%%%%%%%%%%%%%%%%%%%%%%%%%%%%%%%%%%%%%%%%%%%%%%

\section{Contribution of vector mesons
\label{VectorMesonReach}}
\begin{figure}[!tbh]
\centering
\includegraphics[width=0.45\textwidth]{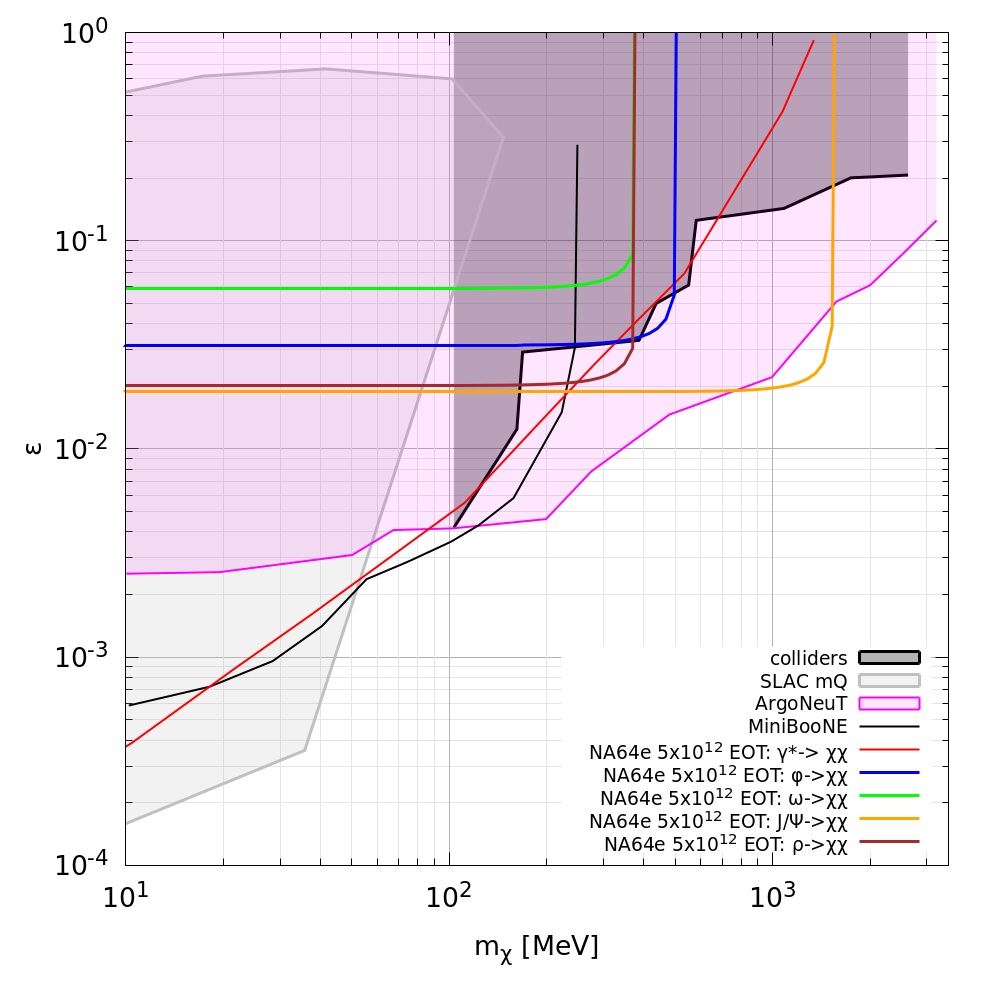}
\caption{The expected sensitivity (90\%\,C.L.) of NA64$e$ in the $(\epsilon, m_\chi)$ plane. We take into 
account invisible decays of vector mesons to the MCP, $V\to \chi \bar{\chi}$, and MCP production by the 
energetic beam electrons via bremsstrahlung-like mode  $\gamma^*\to \chi \bar{\chi}$ for
the prospect statistics $N_{EOT}=5\times 10^{12}$ and MCP mass range 
$10 \, \mbox{MeV}\leq m_{\chi} \leq 1.5$~GeV.  
\label{MesonNA64Limits} }
\end{figure}

In this Section we estimate a contribution of vector mesons, $V=\{ \rho, \omega, \phi, J/\psi \}$, to MCP production at NA64$e$. We adopt the results of 
Ref.~\cite{Schuster:2021mlr} to estimate the yield of vector mesons
produced at NA64$e$ in photo-nuclear reaction $\gamma N \to N V$ for the expected  statistics $N_{EOT}\simeq 5\times 10^{12}$.
The invisible branching ratio of mesons to MCP can be written as

\begin{align}
 & \mbox{Br}(V\to \chi \bar{\chi}) = \epsilon^2 \times \mbox{Br}(V\to e^+e^-)  \times \\
 & \times \left(1+2m_{\chi}^2/m_{V}^2\right) \left(1-4 m_{\chi}^2/m_V^2\right)^{1/2}\,.    \nn 
\end{align}
In the absence of signal events, it implies the limit on $\epsilon$ for each meson mode, 
$V\to \chi \bar{\chi}$, of the following form
\begin{align}
& \epsilon \gtrsim N_{V}^{-1/2} \cdot s_{up}^{1/2} \cdot    \left(\mbox{Br}(V\to e^+e^-) \right)^{-1/2}
\times 
\\
& \times  \left(1+2m_{\chi}^2/m_{V}^2\right)^{-1/2}
\left(1-4 m_{\chi}^2/m_V^2\right)^{-1/4}\,.
\end{align}
Here $N_V$ denotes the total number of $V$-mesons produced at NA64, which is taken from Tab.~II of 
Ref.~\cite{Schuster:2021mlr}.  
In Fig.~\ref{MesonNA64Limits} 
we show the corresponding limits in $(\epsilon, m_{\chi})$ plane. One observes a patch around $m_\chi\simeq 1$\,GeV and $\epsilon\simeq 2\cdot 10^{-2}$ in the allowed region which can be probed at NA64$e$ with MCP from decays of $J/\psi$. 

%%%%%%%%%%%%%%%%%%%%%%%%%%%%%%%%%%%%%%%%%%%%%%%%%%%%%%%%%%%%%%%%%%%%%%%%%%%%

\section{Conclusions}\label{Conclusion}

In the present paper we investigate various effects associated with MCP passage through the matter. In particular,
we discuss in detail the MCP energy losses in matter due to the ionization, 
bremsstrahlung and $e^+e^-$ pair production. We also discuss the effects of the
MCP trajectory deflection due 
to the multiple scattering in the matter. We implement these results to revise the bounds on the MCP parameters 
from SLACmQ experimental data.   We find that our results are in a good agreement with the 
previous study \cite{Prinz:2001qz}. Our study opens a part of the model parameter space previously considered as excluded on the base of published SLACmQ results\,\cite{Prinz:1998ua}. 

In addition, by exploiting the state-of-the-art {\tt CalcHEP} package we calculate 
the exact tree-level cross section of the electron scattering off nucleus $e^- N \to e^- N \gamma^*(\to \chi \bar{\chi})$ to estimate the 
sensitivity of NA64$e$ fixed target experiment to MCP parameters. 
We find that the bremsstrahlung reaction $\chi N 
\to \chi N \gamma$ is the dominant process of MCP energy losses in the detector of NA64$e$ for 
the parameter space of interest $\epsilon \lesssim 0.1$ and $m_\chi \lesssim 1$\, MeV. To summarise, NA64$e$ can test the models with MCP masses from about $10^{-4}$\,eV to 1\,GeV. In particular,  
we show that for  the expected statistics of electrons incident on target $N_{EOT}\simeq5\times 10^{12}$ at NA64$e$
the relatively light MCP  with $m_\chi \simeq  10^{-4}$~eV and $\epsilon \simeq 10^{-5}$
can be directly probed provided by the bremsstrahlung-like missing energy process 
$e^-N \to e^- N \gamma^{*}( \to \chi \bar{\chi})$, while the relatively large MCP with $m_\chi\simeq 1$\,GeV and  
$\epsilon\simeq 2\cdot 10^{-2}$ can be examined thanks to  the invisible vector meson decay signature 
$eN \to e N J/\psi(\to \chi \bar{\chi})$. 

% \vskip 0.5cm
{\it Acknowledgements.}  
We would like to thank R.\, Capdevilla, A.\, Celentano, X.\, Chu, P.\, Crivelli, S.\, Demidov, 
D.\, Forbes, Y.\, Kahn, M.\, Kirsanov, N.\, Krasnikov, 
G.\, Krnjaic, G.\, Lanfranchi, V.\, Lyubovitskij,  
L.\, Molina Bueno, J.\, Pradler,  A.\, Prinz, A.\, Pukhov, G.\, Rubtsov, P.\, Satunin,
P.\, Schuster, H.\, Sieber,  F.\, Tkachov\footnote{Deceased} 
and A.\, Zhevlakov for very helpful discussions and  correspondences. 
The work on the estimate of the expected sensitivity of NA64$e$ is partly supported by 
the  Russian Science Foundation  RSF grant 21-12-00379.
 The work of NA on refining SLACmQ sensitivity was supported by the grant of ''BASIS'' Foundation
 21-2-1-100-1.

% \newpage 

\appendix

%%%%%%%%%%%%%%%%%%%%%%%%%%%%%%%%%%%%%%%%%%%%%%%%%%%%%%%%%%%%
\section{Factors for the spectrum of pair production \label{appendixA}}
The function is defined by
\begin{align}
&     G = -A\, g\left(\dfrac{1}{1+\xi}\right)-B\xi\log{\left(1+\dfrac{1}{\xi}\right)}-\dfrac{C}{1+\xi}+ \nn
\\
&+  \left( 2\log{\dfrac{E_{e^+}\, E_{e^-}}{2m_e\omega\sqrt{1+\xi}}}-1\right)\times \nn 
\\
& \times\left( A\, \log{\left(1+\dfrac{1}{\xi}\right)}+B+\dfrac{C}{1+\xi}\right) 
 \label{eq8} 
\end{align}
where 
$$ 
A=\left(1-\dfrac{4}{3}\dfrac{E_{e^+}\, E_{e^-}}{\omega^2}\right)\left(1+\dfrac{\omega^2}{2 E_\chi E_\chi'}\right)+\dfrac{4}{3}\xi\left(1-\dfrac{E_{e^+}E_{e^-}}{\omega^2}\right)\,,
$$ 
$$
B=\dfrac{4}{3}\dfrac{E_{e^+}E_{e^-}}{\omega^2}-1\;, \quad C=-\dfrac{\xi}{3}-\dfrac{1}{6}\dfrac{\omega^2}{E_\chi E_\chi'}-\dfrac{1}{3}\dfrac{(E_{e^+}-E_{e^-})^2}{\omega^2}\,,
$$
$$
\xi=\dfrac{m_\chi^2 E_{e^+}\, E_{e^-}}{4m_e^2E_\chi (E_\chi-\omega)}\;, \quad g(x)=-\int_0^x\log{\dfrac{|1-t|}{t}}dt\;,
$$
and 
$$
\omega= E_{e^+}+E_{e^-}\,, \;\;\; E_\chi=E_{\chi'}+\omega\,.
$$
%%%%%%%%%%%%%%%%%%%%%%%%%%%%%%%%%%%%%%%%%%%%%%%%%%

\section{Numerical approximation for the energy loss
\label{appendixB}}

This Appendix presents the approximation formulae for the MCP energy loss due to bremsstrahlung and $e^+e^-$ pair production. 
The bremsstrahlung energy losses can be approximated as:
\[
\left|\dfrac{dE_\chi}{dx}\right|_{\text{brems. approx}}\approx0.45\, \mbox{cm}^{-1} E_\chi \epsilon^4 \left(\frac{m_\chi}{ 1\,\mbox{MeV}}\right)^{-2}
\]
The MCP energy loss due to $e^+e^-$ pair production can be described by different expressions for different ranges of MCP mass: 

\begin{align}
&  \left|\dfrac{dE_\chi}{dx}\right|_{\text{$e^+e^-$ pair approx}}= 
\\ 
& 10^{-a_i} \epsilon^2 \left(\dfrac{E_\chi}{ 1\, \text{MeV}}\right)^{d_i} \left(\dfrac{m_\chi}{ 1\,\text{MeV}}\right)^{c_i}\, \left[\text{MeV }\text{cm}^{-1}\right]
\end{align}
where  $e_{i-1}<\dfrac{m_\chi}{\text{MeV}}<e_i,$ and $i=1,\dots,5$,  values of the coefficients are shown below: $e_0=0$, 
\[
\begin{array}{|l|l|l|l|}
\hline a_1=1.9281 & c_1= -0.0805 & d_1=1.1340 & e_1=2.154\times 10^{-4}\\
\hline a_2=2.3194 & c_2= -0.1964 & d_2=1.1374 & e_2=0.1551\\
\hline a_3=2.7185 & c_3= -0.5705 & d_3=1.1477 & e_3=4.1596\\
\hline a_4=2.6936 & c_4= -0.3986 & d_4=1.1138 & e_4=12.4520\\
\hline a_5=2.8549 & c_5= 0.1423 & d_5=1.0096 & e_5=1\times 10^3\\
\hline
\end{array}
\]
Recall Sec.\,\ref{DerivStopLossAndScatt}, that the ionization loss of MCP is negligible for the interesting energetic MCP, $E_\chi>10$\,GeV. 
Thus the approximate overall energy loss of high energy MCP is given by the sum
\[
\left|\dfrac{dE_\chi}{dx}\right|_{\text{rad. approx.}}=\left|\dfrac{dE_\chi}{dx}\right|_{\text{brems. approx.}}+\left|\dfrac{dE_\chi}{dx}\right|_{\text{$e^+e^-$ pair approx.}}
\]

\begin{figure}[!tbh]
\centering
\includegraphics[width=0.45\textwidth]{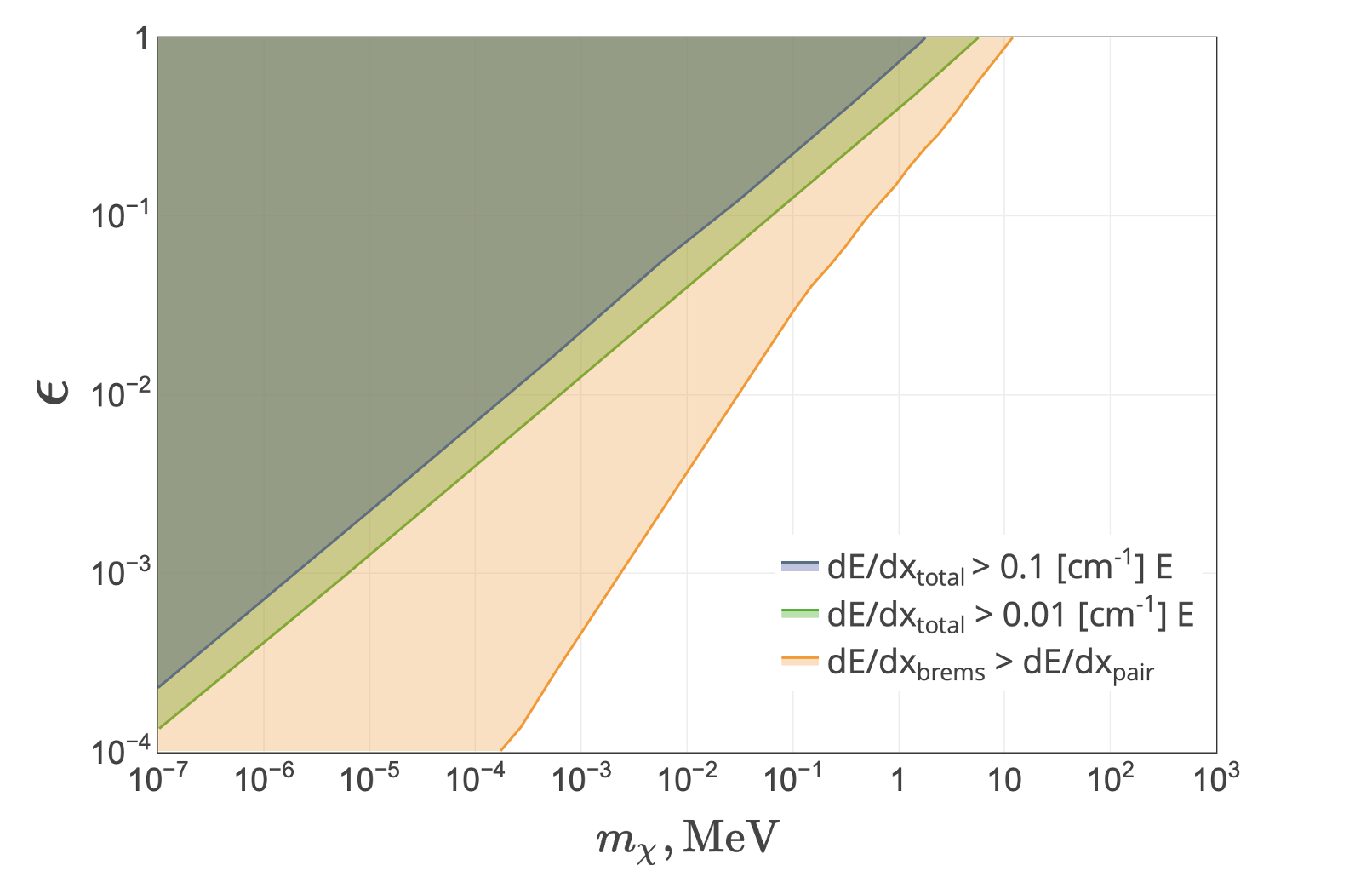}
\caption{The bremsstrahlung energy losses dominate in the orange area. In the regions colored dark green, the energy losses can not be considered as small. 
\label{dominate_brems} }
\end{figure}

%\bibliographystyle{unsrt} 
%\bibliographystyle{alpha} 

%\bibliographystyle{plain} 
%\bibliography{bibl}

%\newpage

\end{document}